\documentclass[%
 aip,
 amsmath,amssymb,
reprint,%
]{revtex4-1}

\usepackage{graphicx}
\usepackage{dcolumn}
\usepackage{bm}

\usepackage[utf8]{inputenc}
\usepackage[T1]{fontenc}
\usepackage{mathptmx}
\usepackage{color}
\usepackage{appendix}
\usepackage{braket}

\usepackage{caption}
\usepackage{subcaption}
\usepackage{comment}

\usepackage[labelsep=period]{caption}
\renewcommand{\thetable}{\arabic{table}}
\usepackage{multirow}
\usepackage{threeparttable}
\usepackage{eucal}

\begin{document}

\preprint{AIP/123-QED}

\title[]{
Applying generalized variational principles to excited-state-specific
complete active space self-consistent field theory
}

\author{Rebecca Hanscam}
\affiliation{
Department of Chemistry, University of California, Berkeley, California 94720, USA 
}

\author{Eric Neuscamman}
\email{eneuscamman@berkeley.edu}
\affiliation{
Department of Chemistry, University of California, Berkeley, California 94720, USA 
}
\affiliation{Chemical Sciences Division, Lawrence Berkeley National Laboratory, Berkeley, CA, 94720, USA}

\date{\today}

\begin{abstract}
  We employ a generalized variational principle to improve the stability, reliability, and precision of fully excited-state-specific complete active space self-consistent field theory. Compared to previous approaches that similarly seek to tailor this ansatz's orbitals and configuration interaction expansion for an individual excited state, we find the present approach to be more resistant to root flipping and better at achieving tight convergence to an energy stationary point. Unlike state-averaging, this approach allows orbital shapes to be optimal for individual excited states, which is especially important for charge transfer states and some doubly excited states. We demonstrate the convergence and state-targeting abilities of this method in LiH, ozone, and MgO, showing in the latter that it is capable of finding three excited state energy stationary points that no previous method has been able to locate.
\end{abstract}

\maketitle

\section{Introduction}

Whether one looks at carotenoids,
\cite{carotenoids-review,carotenoids-1,carotenoids-2}
photochemical isomerization, \cite{PI-1,PI-2,PI-3}
or transition metal oxide diatomics, \cite{TMO-review,TMO-1,TMO-2}
molecular excited states often display wave function characteristics
that go beyond the simplifying assumptions of mean field theory.
From the right perspective, this fact is not that surprising, as it is the
widening of the HOMO-LUMO gap that helps determine ground state
equilibrium geometries and ensure the validity of mean field theory.
Upon excitation, a molecule may be far from the excited state's equilibrium
geometry, and in any case there is no longer the HOMO-LUMO gap to prevent
near-degeneracies between different fillings of the molecular
orbital diagram that may be important for the state under study.
The result is that methods like time-dependent density functional theory
and equation-of-motion coupled cluster theory that perturb around the
mean field limit, while extremely useful in many excited state contexts,
are qualitatively inappropriate in many others.
Instead, methods that explicitly engage with the strongly
multi-configurational nature of these states are called for.
Ideally, these methods would be equally capable for excited states as they
are for ground states, but, as in so many areas of electronic structure
theory, the current reality is that they are not.


For decades, multi-configurational photochemical investigations have been supported by complete
active space self consistent field (CASSCF) theory,\cite{CASSCF_1982,CASSCF_1985-1,CASSCF_1985-2,CASSCF_1987} but the approximations
introduced in its most common incarnations can cause challenges when
treating high-lying states or states with widely varying characters.
In particular, the state averaging (SA) approach -- in which one finds the
orbitals that minimize the average energy of multiple
configuration interaction (CI) roots -- makes the assumption that all
states of interest can be constructed to a similar degree of accuracy with
one shared set of orbitals.\cite{SA-CASSCF}
This approximation offers important advantages and has long been a standard
and successful approach to excited states in CASSCF,\cite{SA-CASSCF_app1,SA-CASSCF_app2,SA-CASSCF_app3,SA-CASSCF_app4,SA-CASSCF_app5,SA-CASSCF_app6}
but it can also create a number of difficulties.
Most obviously, it is less appropriate in cases where different states
require significantly different orbital relaxations, as occurs in molecules
bearing both local and charge transfer (CT) excitations.
Indeed, SA-CASSCF relative energies during nuclear motion on an charge
transfer excitations' surface can be in error by 10 kcal/mol or more.\cite{2step_ABN}
Further, the state averaging method links all of the states together so that
if one state is not well served by the chosen active space and displays a
non-analytic point on its energy surface, all states, even those well-served by
the active space, will show cusps or discontinuities on their energy surfaces.
Finally, because it is only the average energy that is made stationary with respect
to the wave function variables, evaluating nuclear energy gradients for geometry
optimization or dynamics requires solving difficult response equations which are
indeed approximated in some implementations.\cite{SA_nuc_grad-1,SA_nuc_grad-2}
In ground state CASSCF, by contrast, the state's energy is stationary already and
nuclear gradient evaluations are much more straightforward.
So, although state averaging has been and will continue to be a powerful asset to
quantum chemical investigation, there are many reasons why and many settings in
which a fully excited-state-specific CASSCF would be valuable.

Looking at the wider world of excited state theory, there has been remarkable progress
in formulating fully state-specific methods in recent years, which augurs
well for progress in this direction in CASSCF theory.
Examples of this progress include work in variational Monte Carlo,
\cite{VMC-review,VMC-leon,VMC-penalty}
variance-based self-consistent field (SCF) theory, \cite{sigmaSCF-1,sigmaSCF-2}
more robust level shifting approaches in SCF methods, \cite{Eproj}
core spectroscopy, \cite{core_dip1,core_dip2,core_scott1,core_scott2}
perturbation theory, \cite{ESMP2}
and coupled-cluster theory. \cite{ESCC-1,ESCC-2}
Especially relevant to the current study is the
``W$\Gamma$'' approach to state-specific CASSCF (SS-CASSCF), \cite{2step}
in which an approximate variational principle and density matrix information are used to
carefully follow a particular CI root during a two-step optimization that goes back and forth
between orbital relaxation steps and CI diagonalization steps.
The W$\Gamma$ approach proved capable of overcoming root flipping in a wider
variety of situations than readily-available alternatives,
improving CASPT2 energies when compared to state-averaging,
and in making qualitative improvements to some potential energy
surfaces.\cite{2step,2step_ABN}
However, it was unable to locate at least one of the low-lying states of MgO and,
as a method that lacks coupling between orbital and CI variables,
it struggles to tightly converge stationary points.
The method presented here proves more reliable when faced with root flipping and far
superior at tight convergence thanks to its objective function and its
coupling of orbital and CI parameters during optimization.

To understand how these advantages come about, let us turn to discussing recent
progress in the use of quasi-Newton methods to minimize energy-gradient-based
objective functions, which has proven effective in the context of both
the excited state mean field (ESMF)
ansatz \cite{ESMF,GVP,burton2022energy} and
Kohn-Sham $\Delta$SCF. \cite{SGM}
Essentially, the idea is to search for energy saddle points -- which in full CI (FCI)
would be the exact excited states -- by minimizing the norm of the energy gradient
with respect to the variational parameters.
By relying on either an initial guess sufficiently close to the desired
stationary point \cite{SGM} or a generalized variational principle (GVP)
that can use sought-after properties to steer an optimization towards that
stationary point, \cite{GVP} these approaches have proven capable of
achieving full excited-state-specificity while avoiding root flipping
or variational collapse to lower states.
While the work in this direction so far has mostly been focused on weakly correlated
excited states, there is no formal barrier to applying the GVP approach to the
CASSCF ansatz, which is our focus here.

To perform excited-state-specific optimization of the CASSCF ansatz, we will minimize
a GVP containing the square gradient norm by purely quasi-Newton descent,
eschewing CI diagonalization (except in generating a guess)
and more traditional augmented Hessian approaches to orbital rotations.\cite{AH-1,AH-3}
Of course, it may be that a combination of all of these methods
ultimately proves more efficient, as has recently been found for
the ground state,\cite{2ndMCSCF1,2ndMCSCF2,2ndCASSCF}
but in this first combination of CASSCF with a GVP, we stick to pure quasi-Newton
minimization for simplicity, and so our core computational task is to evaluate
gradients of an objective function that contains the square norm of the energy gradient.
Recent work has provided multiple ways forward here.
On the one hand, automatic differentiation arguments guarantee that in most scenarios,
the requisite derivatives can be derived automatically and will have a cost that is
a modest and constant multiple of the energy evaluation cost.\cite{ESMF}
In many cases, this guarantee can motivate the derivation of analytic forms for these
derivatives,\cite{ESMF-analytic} which are often even more efficient in practice, although not
necessarily simple or easy to implement.
As an alternative, Hait and Head-Gordon have presented a clever finite-difference
approach to these derivatives.\cite{SGM}
Although finite difference will incur some error relative to analytic or automatic
differentiation, their study of orbital optimization shows that this error is small
enough that it does not prevent successful convergence to excited state stationary points.
The key benefit of this approach is that it requires only that the energy gradient
itself be available, and so is more convenient to implement.
Although it is possible that a fully analytic formulation of the energy gradient norm
derivatives would improve the rate of quasi-Newton convergence by avoiding
finite difference errors, we for simplicity adopt the finite difference approach
here and find that optimization remains effective even when orbital and CI parameters
are optimized together.
In future, it may be interesting to explore whether more accurate analytic expressions
improve numerical efficiency and whether mixtures with CI and augmented Hessian orbital
optimizers are worthwhile, but already the present approach to combining CASSCF
with an excited state GVP allows us to succeed in situations where previous CASSCF
approaches fail.
\section{Theory}

\subsection{CASSCF Ansatz}

The standard CASSCF ansatz\cite{CASSCF_1982,CASSCF_1985-1,CASSCF_1985-2,CASSCF_1987} has been the foundation for a wide range of CASSCF derived methods,\cite{1stCASSCF-1,1stCASSCF-2,1stCASSCF-3,1stCASSCF-4,1stCASSCF-5,1stCASSCF-6,1stCASSCF-7,2ndCASSCF} and is the formulation used in the approach introduced here. CASSCF methods classify subsets of the molecular orbitals as closed orbitals each occupied by two electrons, active orbitals with varying occupation, and virtual orbitals that are completely unoccupied. The CASSCF wave function is therefore composed of all possible electronic configurations within the active orbitals, defining the active space. The wave function must also account for orbital relaxation effects as while rotations within the active space are described entirely by changes to the configuration (CI) coefficients, the virtual and closed orbitals remain excluded.
While enlarging the active space captures more orbital relaxation effects via the CI expansion, this quickly becomes computationally infeasible for large systems. In addition, the results of a CASSCF calculation are often used as the input for higher-order methods that recover dynamic correlation, which can further limit the size of the chosen active space.
Instead, to relax the orbital descriptions we incorporate an orbital rotation operator in the wave function, such that
\begin{align}
    \ket{\Psi_{\text{CAS}}} = e^{\hat{X}} \sum_I c_I \ket{\phi_I}
    \label{eq:WF}
\end{align}
where $\ket{\phi_I}$ represents a Slater determinant and $c_I$ is the corresponding CI coefficient. The total number of Slater determinants, and thus CI variational parameters forming $\vec{c}$, is determined by the size of the active space.

\begin{figure}
    \includegraphics[scale=0.06]{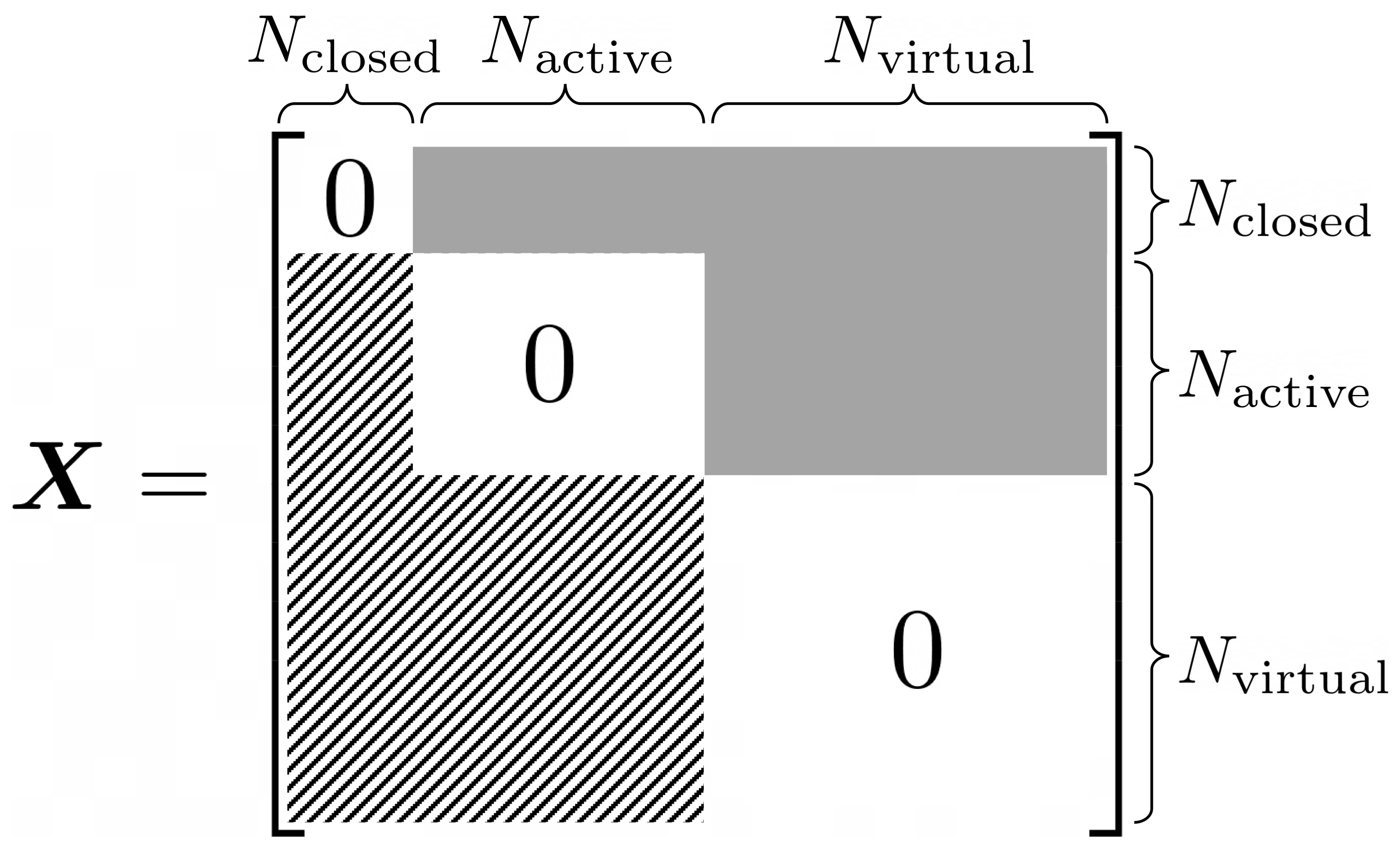}
    \caption{Orbital rotation coefficient matrix $\boldsymbol{X}$ where the solid
    shaded area represents nonzero variational parameters, and the striped region
    is the negative transpose.}
    \label{fig:Xmat}
\end{figure}

For a finite basis of spatial orbitals, the operator $\hat{X}$ in Eq.\ (\ref{eq:WF}) is given by
\begin{align}
    \hat{X} = \sum^{N_{\text{basis}}}_{p<q} X_{pq} \left( \hat{a}^{\dagger}_p \hat{a}_q - \hat{a}^{\dagger}_q \hat{a}_p \right).
    \label{eqn:xdef}
\end{align}
It is defined to be real and spin restricted, thereby ensuring the orbital rotation operator
$\hat{U} = e^{\hat{X}}$ is unitary and also spin restricted. \cite{Helgy,GVP}
Note that only the upper triangle of the matrix $\boldsymbol{X}$ appears
in Eq.\ (\ref{eqn:xdef}), although it is often useful to consider the full matrix,
which is  anti-Hermitian and thus defined by the upper triangle.
Additionally, rotations between orbitals within the active space do not affect the
energy as they are redundant with the flexibility present in the CI expansion.
Similarly, rotations within the closed and virtual orbital spaces have
no affect on the energy.
Were these redundant parameters retained, the variable space would contain
an infinite seam of energetic degeneracy, and so to avoid complications
during numerical optimization, all redundant parameters are excluded.
This choice leads to Figure \ref{fig:Xmat}, which shows the blocks of
$\boldsymbol{X}$ that are included in the orbital variational parameter
set $\vec{x}$.
All together, our CASSCF wave function's variational parameters are the
concatenated set $\vec{v} = \{\vec{c},\vec{x}\}$.

\subsection{Objective Function}
\subsubsection{Generalized Variational Principle}
In FCI, when the energy is expressed as a function of the CI coefficients, the
exact excited states are the energy saddle points of this function.
Even in more approximate theories, the approximate ansatz's saddle points are often
good approximations to the excited states,\cite{deltaSCF,SCF-MOM-1,SCF-MOM-2,ESMF}
and thus the focus of the present investigation is to find excited state
energy stationary points for the CASSCF ansatz.
As these points are not energy minima, gradient-based descent methods are likely
to collapse to lower states, and even non-gradient-based methods like self-consistent
field algorithms can display similar difficulties.\cite{SCF-MOM-1,SCF-MOM-2}
To retain the convenience of minimization algorithms while avoiding this issue
of variational collapse, we choose objective functions that have the square norm
of the energy gradient as their centerpiece.
\begin{align}
    \left| \nabla_{\vec{v}} E \right|^2 = \sum_{i} \left| \frac{\partial E}{\partial c_i} \right|^2 + \hspace{1.2mm} \sum_{j} \left| \frac{\partial E}{\partial x_j} \right|^2
    \label{eq:L_sgm}
\end{align}
In CASSCF, this gradient norm contains contributions from both the CI coefficient 
gradients and the orbital rotation gradients.
It is positive semi-definite by construction, and, for an isolated energy saddle
point, is expected to be surrounded by a basin of convergence that,
if we can somehow get ourselves inside it,
should allow a straightforward minimization of $|\nabla_{\vec{v}} E|^2$
to bring us to the desired excited state energy stationary point.
It is important to note that when $\nabla_{\vec{v}} |\nabla_{\vec{v}} E|^2 =0$
it is possible that $|\nabla_{\vec{v}}E|^2\neq 0$, meaning that the square
gradient norm has stationary points that are not energy stationary points.
In the results discussed below, such cases were overcome through a
combination of improved initial orbital guesses and by incorporating
additional properties within the generalized variational principle
\cite{GVP} (GVP) to which we now turn our attention.

With the norm of the energy gradient being zero for all energy stationary points,
we require some mechanism by which the desired excited state's stationary point
can be targeted.
In some cases, a good enough guess is available to place one within the appropriate
basin of convergence, but in general such a guess may not be available.
To address this problem, we use a GVP approach to expand our objective function
beyond the square gradient norm so that other properties of the excited state can
help steer the optimization into the desired convergence basin.
\begin{align}
    L_{\mu} = \hspace{1.0mm} \mu \left|\vec{d}\hspace{0.5mm}\right|^2
    \hspace{0.5mm} + \hspace{1.0mm}
    (1-\mu) \big|\nabla_{\vec{v}} E \big|^2
    \label{eq:L}
\end{align}
In this objective function, $\vec{d}$ contains functions of the wave function
that should have values close to zero for the desired excited state, such as
the difference $\langle \hat{H} \rangle - \omega$ between the current
wave function energy and a guess for the excited state's energy.
Thus, when $\mu$ is greater than zero and we minimize $L_{\mu}$,
the term containing $\vec{d}$ should help drive the optimization towards
the energy stationary point belonging to the desired excited state.
If the functions within $\vec{d}$ uniquely specify the state (by which
we mean the norm of $\vec{d}$ is smaller for that excited state than
for any other energy stationary point), then an optimization in which
$\mu$ is gradually lowered to zero will arrive at the desired
stationary point. \cite{GVP}

The energy difference term $\langle \hat{H} \rangle - \omega$
that we typically include within $\vec{d}$ can be
motivated as a useful approximation \cite{ESMF,2step}
to the rigorous excited state variational principle
\begin{align}
    W = \frac{\bra{\Psi}(\omega - \hat{H})^2\ket{\Psi}}{\braket{\Psi|\Psi}}
      \approx \big( \langle\hat{H}\rangle - \omega \big)^2,
\end{align}
which if evaluated exactly has its global minimum at the Hamiltonian
eigenstate whose energy is closest to $\omega$. \cite{ESVP-1,ESVP-2}
Of course, many other properties and functions of the wave function can
also be useful in specifying the desired state through the vector $\vec{d}$.
For example, if we knew that it should ideally be orthogonal to
another nearby state $|\Phi\rangle$ and should have a dipole moment
$\vec{\mu}$ (not to be confused with the weighted average parameter
$\mu$ above) of about $\vec{\mu}_0$, we might use
$\vec{d} = \{\langle \hat{H} \rangle - \omega, \hspace{1mm} 
\braket{\Psi|\Phi}, \hspace{1mm} 
| \vec{\mu} - \vec{\mu}_0 | \hspace{1mm}\}$
to guide our optimization into the desired basin of convergence, at
which point $\mu$ can be reduced to zero so that, in the final stage
of optimization, minimization of the energy gradient square norm
brings us to the desired stationary point.
It is important to recognize that the functions employed within $\vec{d}$
need not be exact, as their only purpose is to get us into the right
basin of convergence, after which they have no further effect.
A good example of where this flexibility can be exploited is seen
in our results on ozone, where we use a simple approximation for the
overlap with another state to help one of our
optimizations converge correctly.
Evaluating that overlap exactly would be an exercise in
non-orthogonal CI (NOCI),\cite{noci-1,noci-2,noci-ES} 
but in this case a simple dot product between CI vectors (which neglects
differences in the molecular orbitals) is free by comparison and a good
enough nudge to guide the optimization to the desired stationary point
in the face of a tricky near-degeneracy.

\subsubsection{Objective Function Gradient}

To minimize our objective function via gradient descent, we will need an expression
for its gradient.
When $\vec{d} = \{\braket{\hat{H}}-\omega\}$, this gradient is
\begin{equation}
\begin{aligned}
   \nabla_{\vec{v}} L_{\mu} \hspace{0.5mm} =& \hspace{1.5mm}
   2 \mu (E - \omega) \nabla_{\vec{v}} E
   \hspace{1.0mm} + \hspace{1.0mm}
   (1-\mu) \nabla_{\vec{v}}|\nabla_{\vec{v}} E|^2.
   \label{eq:dL} 
\end{aligned}
\end{equation}
In CASSCF, the energy gradient with respect to the full variational parameter
set $\nabla_{\vec{v}}E$ can be split into the energy gradient with respect to
the CI parameters $\nabla_{\vec{c}}E$ and the energy gradient with respect
to the orbital rotation parameters $\nabla_{\vec{x}}E$.
In this work, we use the analytic expression for the CI gradient
\begin{align}
    \nabla_{\vec{c}} E = \frac{\partial E}{\partial \vec{c}} = \frac{2(H-E)\vec{c}}{\vec{c}^T \cdot \vec{c}}
    \label{eq:dEdC}
\end{align}
where $H$ is the Hamiltonian matrix in the CI basis.
For the orbital energy gradient, we use the analytic expressions given in the SI that are comprised of contractions between the MO integrals and the one and two-electron spin-summed reduced density matrices.
\cite{ESMF-analytic,sigmaSCF-1,SGM,CASSCF_1985-1,CASSCF_1987,DMRG-1}
These expressions assume we are working within the current MO basis (i.e. when $X=0$), the implications of which are discussed in Section \ref{section:objf_approxHess}.

By far the most computationally challenging term in Eq.\ (\ref{eq:dL}) is the derivative of the squared norm of the energy gradient with respect to the variational parameters,
\begin{align}
    &\frac{\partial}{\partial v_j} \left| \nabla_{\vec{v}} E \right|^2
    \hspace{1mm} = \hspace{1mm}
    \frac{\partial}{\partial v_j} \sum_i
    \left| \frac{\partial E}{\partial v_i} \right|^2
    \hspace{0.1mm} = \hspace{1mm} 2 \sum_i \mathcal{H}_{ij}
    \frac{\partial E}{\partial v_i}.
    \label{eq:dEnsq_full}
\end{align}
The Hessian matrix of energy second derivatives
$\mathcal{H}_{ij} \equiv \frac{\partial^2 E}{\partial v_i \partial v_j}$
is expensive to evaluate, and we certainly do not wish to construct it explicitly.
While it is possible to use automatic differentiation to evaluate this term,
\cite{ESMF}
for ease of implementation we instead turn to a central finite difference
method that Hait and Head-Gordon have shown to be effective
for excited state orbital optimization. \cite{SGM}
Using a directional finite difference of the energy gradient with a chosen
perturbation of
$\delta \vec{v}=\lambda \nabla_{\vec{v}}E \big|_{\vec{v}=\vec{v_0}}$
yields the approximate expression
\begin{equation}
    \begin{aligned}
        \nabla_{\vec{v}} \left| \nabla_{\vec{v}} E \right|^2 &= \frac{1}{\lambda} \left( \nabla_{\vec{v}}E \big|_{\vec{v}=\vec{v_0}+\delta \vec{v}} - \nabla_{\vec{v}}E \big|_{\vec{v}=\vec{v_0}-\delta \vec{v}} \right) \\
        &\hspace{5mm}+ O \left( \lambda^2 \left( \nabla_{\vec{v}} E \big|_{\vec{v}=\vec{v_0}} \right)^3   \right).
    \end{aligned}
    \label{eq:fd}
\end{equation}
This approach avoids the computationally demanding Hessian-gradient contraction
in Eq.\ (\ref{eq:dEnsq_full}), replacing it with multiple evaluations of the
energy gradient.
Automatic differentiation -- as its cost is typically 2-3 times the cost of
the function -- should be able to deliver a fully analytic version of this
approach with zero finite difference error at a similar price, as has
been achieved for ESMF.
Further, a hand-implemented analytic version could be even faster.
Thus, it may be worth investigating in future whether the removal of the small
finite difference error leads to a significant improvement in optimization
efficiency.
For the present study, however, we employ Eq.\ (\ref{eq:fd}) as is for both
the orbital and CI variables together and find that it is sufficient
for achieving tight energy stationary point convergence.
It is important to stress that, regardless of which of these
approaches is taken for evaluating the objective function gradient,
the computational cost of doing so is at worst equal to a handful
of CASSCF energy gradient evaluations, and so the scaling of the
approach with system size is the same as in standard CASSCF.

A close inspection of Eq.\ (\ref{eq:dEnsq_full}) shows that, even if one
applies naive steepest descent for minimizing the objective function,
some coupling between the orbital and CI variables is present due to
the energy Hessian.
In practice, a quasi-Newton approach that builds up an approximation to the
objective function Hessian will account for even more coupling between these
variable sets.
Although it is too early to tell how well this approach to coupling works
as compared to second-order ground state approaches, \cite{2ndMCSCF1,2ndCASSCF}
a quasi-Newton minimization of our objective function certainly incorporates
more coupling than a simple two-step optimization \cite{2step}
in which one goes back and forth between optimizing the CI variables
with the orbitals held fixed and optimizing the orbitals with the
CI variables held fixed.
In each step of quasi-Newton minimization, the effects of orbital changes
on the CI energy gradient and CI changes on the orbital energy gradient
are taken approximately into account.
The result is a dramatic improvement in the method's ability to tightly
converge the energy gradient as compared to the two-step W$\Gamma$
approach that we compare to in our results below.

\subsubsection{Approximate Objective Function Hessian}
\label{section:objf_approxHess}

In this work, we use the limited-memory Broyden-Fletcher-Goldfarb-Shanno (L-BFGS)
algorithm \cite{bfgs1,bfgs2,bfgs3,bfgs4} to minimize the objective function.
Roughly speaking, L-BFGS takes a Newton-like step using an approximate Hessian.
In particular, this approximate Hessian is arrived at by using
finite-differences between previous iterations' objective function gradients
to improve upon some initial guess for the objective function Hessian.
This initial guess can be set to the identity matrix for simplicity, but
the speed of convergence can be accelerated dramatically if a better
guess for the Hessian is supplied\cite{preconditionedBFGS}, as has been demonstrated for objective
functions like ours in both the $\Delta$SCF \cite{deltaSCF} and
ESMF \cite{BVDG_thesis} contexts.
Indeed, our approach here is another example  of using a quasi-Newton
method to further improve a CASSCF approximate Hessian scheme.
An early example of using quasi-Newton methods for this purpose
occurred in the context of improving super-CI methodology for
restricted active space wave functions, \cite{RASSCF-unitary}
and very recent work has shown that orbital-CI coupling for
ground state optimizations can be usefully accelerated via
quasi-Newton as well. \cite{2ndMCSCF1,2ndMCSCF2}
In the present study, we see that even if L-BFGS starts from
the identity matrix as the initial Hessian guess, it is better
at achieving tight convergence than an uncoupled
two-step optimization like the $W\Gamma$ method.
However, the smarter approach \cite{RASSCF-unitary,2ndMCSCF1,2ndMCSCF2}
of using a quasi-Newton method like L-BFGS to improve
on a more accurate (although still approximate)
initial Hessian is more effective still, and so we will seed L-BFGS
with diagonal approximations to our objective function's Hessian.

Starting with the Hessian of the $\mu=0$ objective function,
(i.e.\ the second derivatives of the energy gradient square norm)
\begin{equation}
    \begin{aligned}
        \frac{\partial^2}{\partial v_j \partial v_k}
        \sum_{i} \left| \frac{\partial E}{\partial v_i} \right|^2 = 
        \hspace{1mm}2 & \sum_{i} \mathcal{H}_{ij} \mathcal{H}_{ik}\\
        &+ 2 \sum_{i} \left( \frac{\partial^3 E}{\partial v_i \partial v_j \partial v_k} \right) 
        \frac{\partial E}{\partial v_i},
        \label{d2Ensq}
    \end{aligned}
\end{equation}
we can anticipate that, due to its contraction with the energy gradient,
the role of the third derivative tensor will become negligible as
the optimization approaches an energy stationary point.
Indeed, it has been observed empirically in both 
$\Delta$SCF \cite{deltaSCF} and ESMF \cite{BVDG_thesis} that dropping
this term entirely does not much matter, and so we neglect it here as well.
In the case where $\vec{d} = \{\braket{\hat{H}}-\omega\}$ and we now
allow $\mu$ to be zero or nonzero, this leaves us with the following
approximate expression for the objective function Hessian.
\begin{equation}
    \begin{aligned}
        \frac{\partial^2 L_{\mu}}{\partial v_j \partial v_k}
        \approx
          \hspace{1mm} 2 \mu \left[
          (E-\omega) \mathcal{H}_{jk}
          + \frac{\partial E}{\partial v_j}
            \frac{\partial E}{\partial v_k}
        \right] \\
        \qquad
        + 2 (1-\mu) \sum_{i} \mathcal{H}_{ij} \mathcal{H}_{ik}
        \label{eq:d2L}
    \end{aligned}
\end{equation}
When not using the identity, we will use the diagonal of Eq.\ (\ref{eq:d2L})
as the approximate objective function Hessian that we supply to L-BFGS. However, evaluating the full energy Hessian $\mathcal{H}$ is impractically expensive. To make this approach affordable, we extend the diagonal approximation to $\mathcal{H}$ as well, leaving us with the following expression. \begin{equation}
    \begin{aligned}
        \frac{\partial^2 L_{\mu}}{\partial v_i^2}
        \approx
          \hspace{1mm} 2 \mu \left[
          (E-\omega) \mathcal{H}_{ii}
          + \left| \frac{\partial E}{\partial v_i} \right|^2
        \right] \\
        \qquad
        + 2   (1-\mu) \mathcal{H}_{ii}^2
        \label{eq:d2L_diag}
    \end{aligned}
\end{equation}

We approximate the energy Hessian $\mathcal{H}$ in Eq.\ (\ref{eq:d2L_diag}) using a diagonal form, although we make
different choices for how to deal with
the CI block (denoted ${}^{cc}\mathcal{H}$)
and the orbital block (denoted ${}^{xx}\mathcal{H}$).
In the CI block, we make no approximation beyond omitting the off-diagonal
terms, leaving us with the same diagonal that is used in the Davidson
algorithm. \cite{Davidson}
\begin{align}
    {}^{cc}\mathcal{H}_{ii} =
    \frac{2 \left( H_{ii} - E \right) }{\vec{c} \cdot \vec{c}}
\end{align}
For the diagonal of the orbital block, we define
$E_{pq}^- =
\left( \hat{a}_{p}^{\dag} \hat{a}_{q} - \hat{a}_{q}^{\dag} \hat{a}_{p} \right)$
and arrive at the following expression. \cite{Helgy}
\begin{align}
    {}^{xx}\mathcal{H}_{pq,pq} &=
    \frac{\partial^2 E}{\partial x_{pq} \partial x_{pq}} =
    \bra{\Psi} \left[ E_{pq}^- , \left[ E_{pq}^- , \hat{H} \right] \right] \ket{\Psi}
     \label{eq:ooHess}
\end{align}
Following the derivation by Siegbahn \textit{et al.} of the full orbital-orbital energy Hessian using Fock-like matrices,\cite{orbHess,1stCASSCF-6} explicit expressions for the exact diagonal of ${}^{xx}\mathcal{H}$ in terms of two-electron integrals and density matrices are provided in the SI for the reader's convenience and have been extensively checked with finite difference. Previous approaches in second-order MCSCF methods make further approximations to the diagonal of ${}^{xx}\mathcal{H}$, demonstrating this to be sufficient to achieve improved convergence.\cite{orbHess_approx,1stCASSCF-2} In addition to implementing the exact diagonal expressions and unlike the CI block, we go beyond just dropping the off-diagonal terms by approximating the Hamiltonian inside the commutators with the one-electron Fock operator built from our CASSCF wave function's one-body density matrix. These choices for our approximate energy Hessian diagonal, which are similar to those made in other contexts, \cite{deltaSCF,BVDG_thesis} combine with Eq.\ (\ref{eq:d2L_diag}) to provide L-BFGS with a much better guess than the identity for the objective function Hessian. The Fock-based diagonal improved guess comes at an additional computational cost that is significantly less than the energy gradient evaluation we are already doing, as it involves no two-electron AO-to-MO integral transforms and has a much simpler interaction with the CI vector. While the exact diagonal version necessitates additional AO-to-MO integral transforms not already performed, for the small molecules considered in this study we find the increased cost to be off-set by the convergence speed-up it offers.


In practice, the working equations for the gradients and Hessian elements
we need are simpler when the orbital rotation matrix $X$ is equal to
zero, as it is at the start of the optimization.
However, if one uses the straightforward parameterization of the
$i$th iteration's molecular orbitals as a single rotation from the
initial guess,
\begin{align}
    C_i = C_0 e^{X} 
\end{align}
then at all iterations aside from the first, one must deal with
a non-zero $X$ matrix.
If, instead, one resets the definition of the molecular
orbitals so that $X$ becomes
the rotation from the previous iteration's orbitals
\begin{align}
    C_i = \tilde{C} e^{X}
        = C_0 e^{ X_1} e^{ X_2}... e^{ X_{i-1}}  e^{X} 
\end{align}
then the working equations at each iteration enjoy the simplicity
offered by having $X=0$.
However, when we reset the definition of $X$ in this way, we cause the gradient
history we have accrued to no longer be quite correct, as those
gradients were evaluated with a slightly different definition of
the variables.
In previous work on single-determinant wave functions, \cite{GDM}
it has been shown that the gradient history can be exactly corrected
to account for this change of variables.
For simplicity, we have not done so here, and this has not prevented
our approach from achieving tight convergence for excited states.
However, making these types of gradient history corrections will
presumably accelerate our rate of convergence, and so we look forward
to investigating these corrections in future efforts to improve
numerical efficiency, which could also benefit from
the use of more sophisticated initial Hessians with non-zero
orbital-CI blocks.


\subsection{Optimization Procedure}
\label{section:opt}
The overall quasi-Newton optimization procedure for our GVP approach to
excited state CASSCF is as follows.
\begin{enumerate}
    \item An initial orbital basis and active space are chosen and an initial guess for the CI coefficients is selected, typically taken from a CASCI calculation or an initial SA-CASSCF calculation. The orbital rotation coefficients are initialized as zero and a value for $\omega$ is estimated from the energy of the initial inputs, results from other methods, or experimental data. 
    \item The set of variational parameters $\vec{v}=\{\vec{c},\vec{x}\}$ are optimized all together via a series of L-BFGS minimizations of $L_{\mu}$ for decreasing values of $\mu$.  We supply either the identity or an approximate objective function Hessian discussed in the previous section as the initial guess for the L-BFGS Hessian.  The initial $\mu$ value and convergence threshold are set to $0.5$ and $|\nabla_{\vec{v}}L|=10^{-3}$, respectively.  Within each micro-iteration of an L-BFGS minimization, the following tasks are completed.
    \begin{enumerate}
        \item The gradient of the objective function $\nabla_{v}L_{\mu}$ with respect to the CI coefficients $\vec{c}$ is built from the analytical expression in Eq.\ (\ref{eq:dEdC}) where the contraction of the active space Hamiltonian with the CI coefficient vector is performed utilizing PySCF's\cite{PYSCF} existing direct CI functions.
        \item The gradient with respect to the orbital rotation coefficients $\vec{x}$ evaluated at $X=0$ is built from Eq.\ (\ref{eq:oGrad_1})-(\ref{eq:oGrad_3}). The scaling of this task is dominated by the AO-to-MO integral transformations.
        \item The value of the finite difference $\lambda$ is set to the maximum of $\{10^{-6}, |\nabla_{\vec{v}}E|\}$ at each iteration, and the objective function (Eq.\ (\ref{eq:L})) and its gradient (Eq.\ (\ref{eq:dL})) are built at the cost of three gradient evaluations of both $\nabla_{\vec{c}}E$ and $\nabla_{\vec{x}}E$.
        \item If the approximate objective function Hessian (Eq.\ (\ref{eq:d2L_diag})) is in use, then it is built using either the exact energy Hessian diagonal or its Fock-based approximation as discussed in the previous section.
        \item Take the L-BFGS step and, afterwards, update the
        definition of the MOs as discussed in the previous section so that
        $X=0$ again.
    \end{enumerate}
    \item After each L-BFGS minimization (macro-iteration), we reduce $\mu$.
    If the maximum element of $|\nabla_{\vec{v}}E|$ is now less than the
    current convergence threshold, then we jump to the final optimization stage,
    setting $\mu=0$ and the convergence threshold to its final value
    of $10^{-7}$ and repeating step 2.
    Otherwise, we decrease $\mu$ by $0.1$ and tighten the convergence threshold
    by a factor of 10 (if it is not yet $10^{-7}$) and repeat step 2.
\end{enumerate}

\section{Results and Discussion}

In the following collection of molecular examples, we aim to answer the key question of how does the GVP approach compare to other SS-CASSCF methods? Is the GVP able to find the CASSCF energy stationary point that corresponds to the initial CASCI root in the face of root-flipping? How does the convergence of the GVP approach compare to other SS-CASSCF methods, with and without the approximate diagonal Hessian being provided to L-BFGS?
Finally, are there situations where the GVP can succeed when other SS-CASSCF methods fail?

These questions were investigated in LiH, asymmetrically stretched
O\textsubscript{3}, and MgO.
The cc-pVDZ atomic orbital basis\cite{ccpVDZ-1,ccpVDZ-2}
was used throughout.
Both LiH and O\textsubscript{3} used the HF orbital basis for the initial guess,
while MgO used the local density approximation (LDA) orbital basis.
An initial CASCI calculation was performed for each of these molecules
and the targeted root's CASCI CI vector was used as the initial guess for
the CI coefficients.
Values for $\omega$ were chosen using past results from other CASSCF
calculations or estimated based on the initial CASCI energy orderings.
The first macro-iteration of each GVP optimization performed in this study held
the CI parameters fixed while converging the orbital gradient to
$|\nabla_{\vec{x}}L|< 10^{-5}$, using the identity as the objective
function Hessian guess.
Beyond the first macro-iteration, all parameters were optimized together
with the approximate diagonal Hessian guess -- built from the exact diagonal energy Hessian -- employed for all values
of $\mu$ in all optimizations in LiH, O\textsubscript{3} and MgO.

In this study, we consider a stationary point converged in our GVP
optimization when
$\left|\nabla_{\vec{v}}|\nabla_{\vec{v}}E|^2\right| < 10^{-7}$,
$|\nabla_{\vec{c}}E|< 10^{-6}$, and $|\nabla_{\vec{x}}E|< 10^{-6}$.
For each of the molecules in this study, the results of the GVP approach are
compared to those of the $W\Gamma$ and
simple root selection (SRS) 2-step methods.
In SRS, one selects the CI root to use in orbital optimization by
always taking the $n$th root from the energy-ordered CI roots, whereas
$W\Gamma$ uses an approximate variational principle and the one-body
density matrix to select the desired root. \cite{2step}
For both $W\Gamma$ and SRS, neither of which has orbital-CI coupling
in our implementation, we set looser convergence thresholds because
this lack of coupling prevents them from converging to the
same level of precision.
For the change in energy, the norm of the orbital gradient, and the
norm of the change in the one-electron density matrix, the $W\Gamma$ thresholds were set to
$10^{-7}$, $10^{-4}$, and $10^{-4}$ respectively.
To check whether a loosely converged $W\Gamma$ or SRS calculation
corresponds to the same stationary point as the GVP, we have therefore
also used our GVP approach to finalize their convergence.
This finalization was never observed to alter the character of the wave function,
even in cases where a non-negligible energy change was observed during finalization.
All molecular orbital analysis was performed with the programs
Gabedit \cite{gabedit} and Molden. \cite{molden:2017}

\subsection{LiH}

The ground state of LiH ($X^1 \Sigma^+ $) is ionic at it's equilibrium
bond length of $1.8$ \text{\AA}, but the first excited state
($A^1\Sigma^+$) is mostly neutral due to a HOMO-LUMO
charge transfer excitation.
However, as the bond is stretched, the ground state becomes increasingly
neutral while the first excited state becomes more ionic.
What makes this an especially interesting molecule to study in the
present context is the avoided crossing that exists between the ground
and first excited states at intermediate bond lengths. \cite{LiH,LiH-2}
The mixing of state characters in this region leads to a well known
root flipping problem \cite{2step,GVP,LiH,LiH-2,SA-CASSCF,2ndMCSCF-SR}
that provides a good test for our GVP approach.

\begin{figure}
    \begin{subfigure}{.5\textwidth}
        \centering
        \includegraphics[scale=0.128]{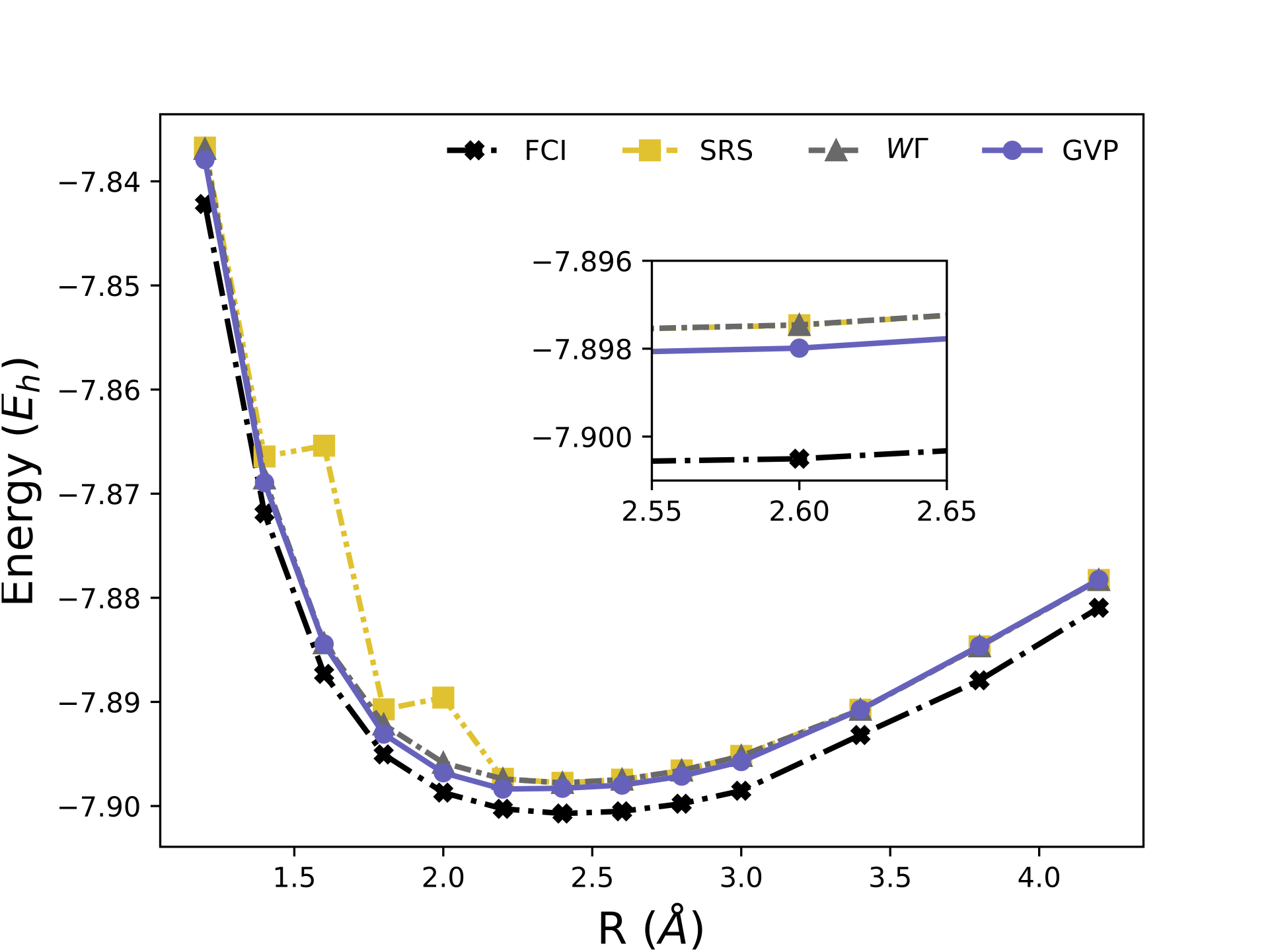}
    \end{subfigure}
    \begin{subfigure}{.5\textwidth}
        \raggedright
        \includegraphics[scale=0.40]{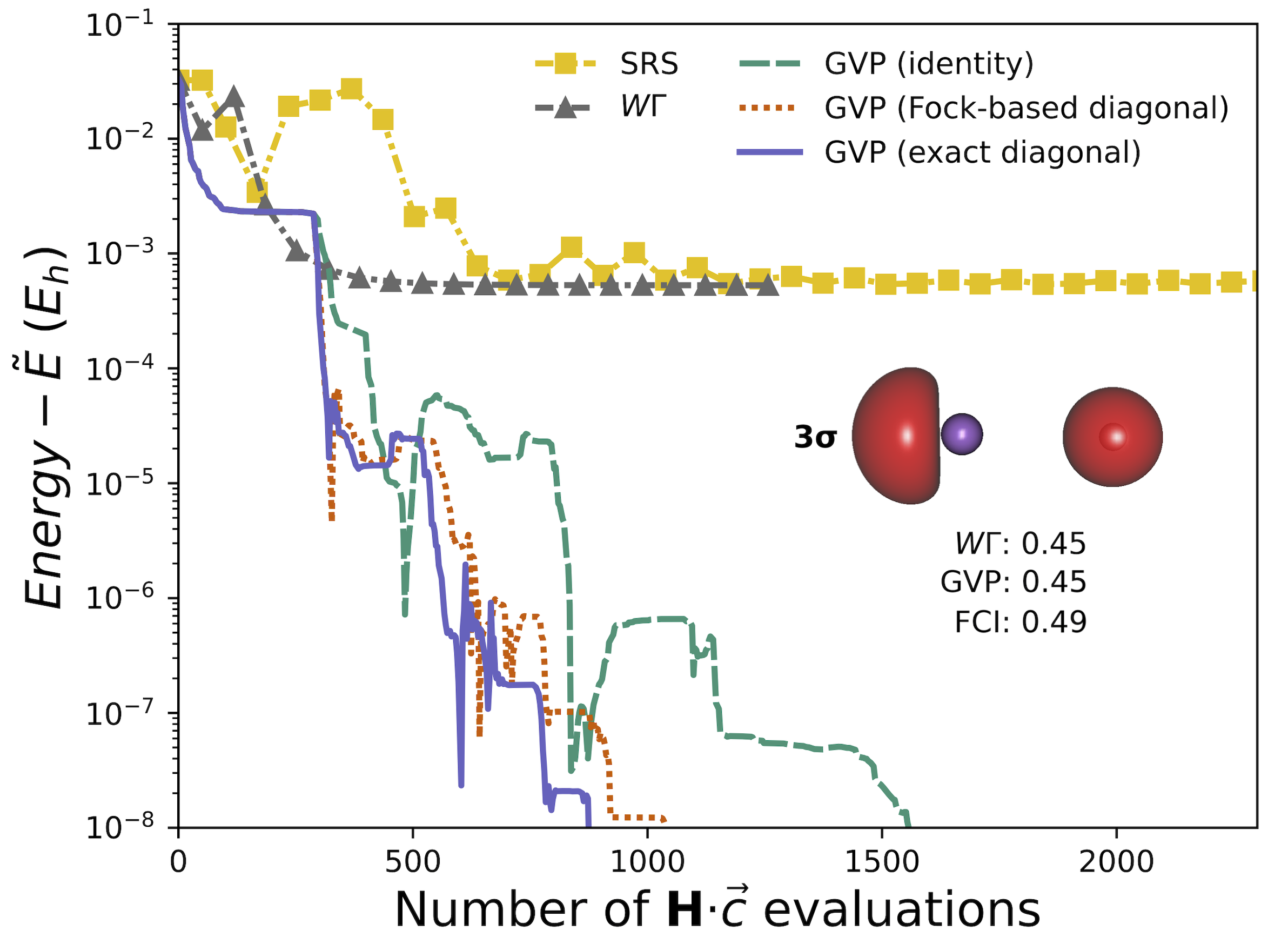}
    \end{subfigure}
    \begin{subfigure}{.5\textwidth}
        \raggedright
        \includegraphics[scale=0.40]{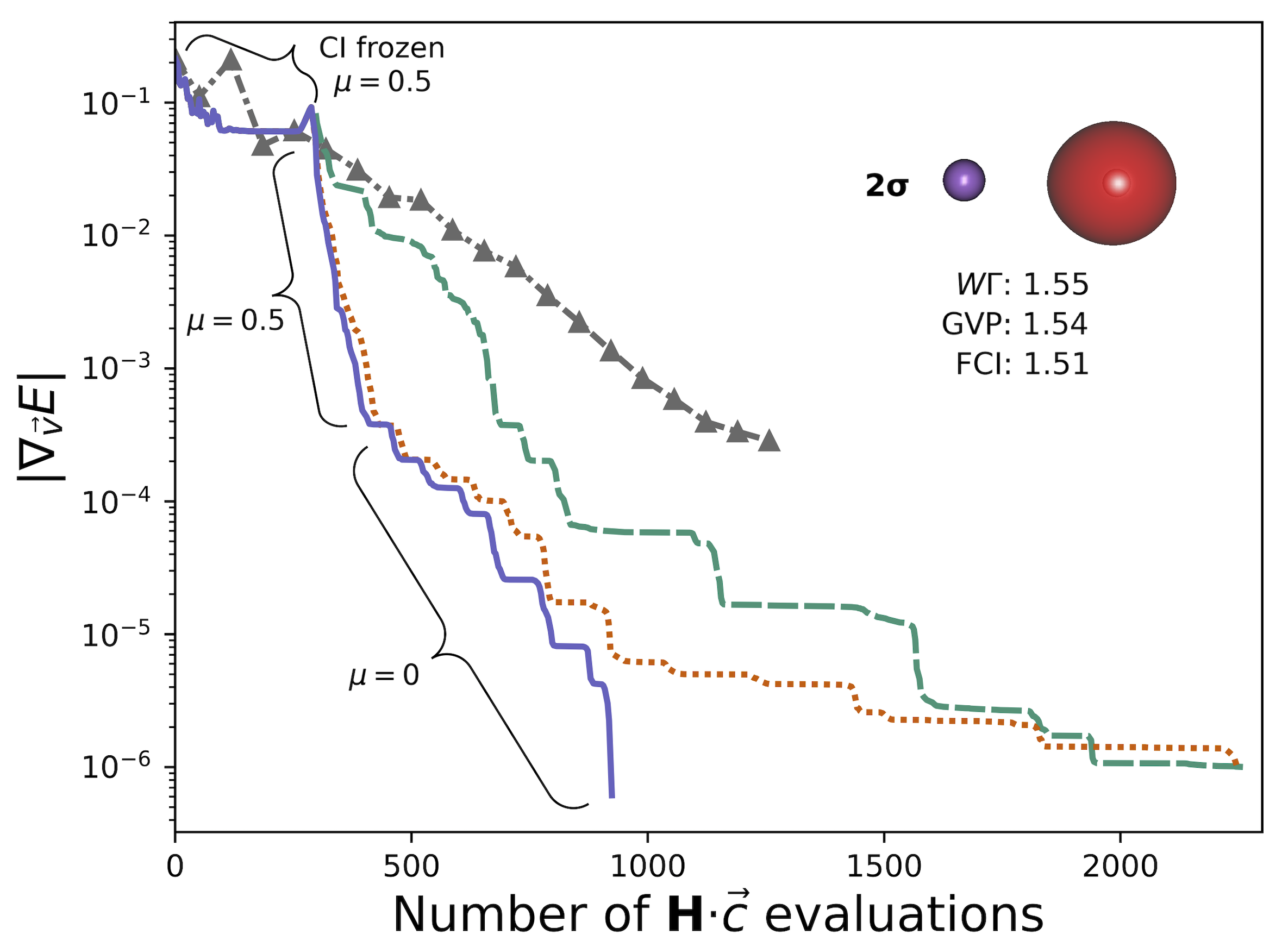}
    \end{subfigure}
    \caption{The top panel shows potential energy surfaces for
    the first excited state of LiH.
    The middle panel shows energy convergence at a bond length of $2.6$ \text{\AA}
    relative to the GVP's final tightly converged energy $\tilde{E}$.
    The bottom panel shows, again at $2.6$ \text{\AA},
    the convergence of the norm of the energy gradient.
    In the middle and bottom panels, the optimization details are labeled for each
    macro-iteration of the GVP approach employing the $\mu$ update schedule as described in Section \ref{section:opt}, with $\omega = -7.9$ E$_h$ used at all macro-iterations. Convergence of the GVP is shown using the identity as the initial Hessian guess (dashed green line), compared to an approximate Hessian built from the exact diagonal (solid purple line) or Fock-based approximate diagonal (dotted orange line) energy Hessian.
    The insets to the middle and bottom panels show the $2\sigma$ and $3\sigma$ natural orbitals and corresponding occupation numbers.
    At each geometry, SRS and $W\Gamma$ converged the orbital gradient to $10^{-4}$,
    while the GVP converged to $10^{-7}$.}
    \label{fig:LiH_energy}
\end{figure}

Using an active space of 4 electrons in 4 orbitals (Li 1s2s2p\textsubscript{z}, H 1s), Figure \ref{fig:LiH_energy} demonstrates that SRS clearly suffers from the root flipping problem, causing it to struggle with convergence and taking a comparatively large number of iterations or failing altogether. Past work\cite{2step} has shown that the $W\Gamma$ method is able to overcome the root flipping problem by tracking the targeted root through the optimization, producing the smooth potential energy surface seen in the top panel of Figure \ref{fig:LiH_energy}. While the dissociation curves illustrate the agreement between $W\Gamma$ and the GVP approaches across all geometries, they also highlight the improvement the GVP achieves in overall convergence, in particular the magnitude of the orbital gradients, by several orders of magnitude from both the SRS and $W\Gamma$ results. For a geometry of $2.6$ \text{\AA}, Table \ref{tab:LiH_wfchar} shows very similar wave function character between the energy stationary point the GVP finds and the more loosely converged $W\Gamma$ state.
Both have strong overlap to the initial CASCI root and it is clear they are both describing the desired state, one is merely more tightly converged than the other.
Indeed, looking at the convergence for this geometry in the bottom panels of Figure \ref{fig:LiH_energy}, the GVP achieves an energy half a mE$_h$ closer to the FCI result than the other state-specific methods in fewer
Hamiltonian-CI-vector multiplies when using an approximate initial
Hessian guess in L-BFGS.
It is especially noteworthy that when using the identity as
the initial Hessian guess we still take a comparable number of Hamiltonian-CI vector contractions, suggesting that helpful orbital-CI coupling is
indeed present in the quasi-Newton approach even without the better Hessian starting guess.

\begin{table}[htbp]
\caption{Wavefunction character in the CASCI orbital basis of the first excited state $A^1\Sigma^+$ of LiH at a bond length of 2.6 \AA.}
\small
\begin{tabular}{r c l c | c c c }
\hline\hline
 & & & Active & & & \\ 
\multicolumn{3}{c}{Primary}     & Space Electron & \multicolumn{3}{c}{\underline{Wavefunction Weight (\%)}} \\
\multicolumn{3}{c}{Excitations} & Configuration         & CASCI  &  $W\Gamma$ & GVP    \\ \hline
2$\sigma$   & $\rightarrow$ & 3$\sigma$            & 1$\sigma^2$ 2$\sigma$ 3$\sigma$ & 86.5 & 82.5 & 82.9 \\ 
2$\sigma^2$ & $\rightarrow$ & 3$\sigma^2$          & 1$\sigma^2$ 3$\sigma^2$         & 5.3  & 5.7  & 5.7 \\ 
2$\sigma^2$ & $\rightarrow$ & 3$\sigma$, 4$\sigma$ & 1$\sigma^2$ 3$\sigma$ 4$\sigma$ & 4.2  & 5.6  & 5.6 \\ 
\multicolumn{3}{c}{Aufbau}                         & 1$\sigma^2$ 2$\sigma^2$         & 3.2  & 5.3  & 4.9 \\ \hline
\multicolumn{4}{c}{Overlap with CASCI Root:} & 1 &  0.95 & 0.96 \\ \hline\hline
\end{tabular}

\label{tab:LiH_wfchar}
\end{table}

\subsection{Asymmetrical O\textsubscript{3}}

We turn next to asymmetrically stretched ozone, which contains two excited states
that are close to energetically degenerate and
prove to be especially challenging for the GVP approach.
Indeed, at this particular geometry
($R_{O_{1}O_{2}} = 1.3$ \text{\AA}, $R_{O_{2}O_{3}} = 1.8$ \text{\AA},
$\angle O_{1}O_{2}O_{3}=120$\textdegree{}),
the $4^{1}A"$ and $5^{1}A"$ states can switch order with each other and
even strongly re-mix their primary configurations depending on the size of
the active space used and whether or not the orbitals are optimized
state-specifically.
We employ a 9-orbital, 12-electron active space and freeze the electronic occupation and orbital shapes of the six lower
energy orbitals (which are, roughly speaking, the O 1s and 2s orbitals).
With this choice, we do in fact observe a root flip: SS-CASSCF optimizations
starting from the 4th and 5th ${}^{1}A"$ CASCI roots find two
different energy stationary points, but the stationary point found when
starting from the 5th CASCI root (and which is most similar in character
to the 5th CASCI root) has a lower energy than the other stationary point,
as displayed in Table \ref{tab:O3_wfchar}.
\begin{table*}[htbp]
\caption{Wavefunction data in the CASCI orbital basis for the 4th and 5th ${}^1$A" states in O\textsubscript{3}.
         Note that the 5th CASCI root ultimately optimizes to become the $4^1$A" state,
         and so its data is presented under the $4^1$A" heading in the left column,
         whereas the 4th CASCI root's data is presented on the right under the $5^1$A" heading.
         The GVP data are for the stationary point found when starting from the
         CASCI root shown under the same heading.
}
\begin{tabular}{r c l| c c | c c}
\hline\hline
 & & & \multicolumn{2}{c|}{\underline{$4^1$A" Wavefunction Weight (\%)}} & \multicolumn{2}{c}{\underline{$5^1$A" Wavefunction Weight (\%)}}\\
\multicolumn{3}{c|}{Primary Excitations} & CASCI & GVP & CASCI & GVP\\ \hline
\ \ \ \ \ 9a', 10a' & $\rightarrow$ & 3a", 11a' & 67.3 & 65.7 & 5.4 & 19.6 \\
\ \ \ \ \ 2a" & $\rightarrow$ & 11a' & 1.8 & 6.0 & 41.0 & 40.1 \\
\ \ \ \ \ 9a', 2a"  & $\rightarrow$ & 3a"$^2$ & 1.4 & 0.0 & 10.0 & 4.0 \\\hline
\multicolumn{3}{c|}{Overlap with 4th ${}^1$A" CASCI root} & 0 & 0.41 & 1 & 0.66 \\
\multicolumn{3}{c|}{Overlap with 5th ${}^1$A" CASCI root} & 1 & 0.87 & 0 & 0.68 \\ \hline
\multicolumn{3}{c|}{Energy (E$_h$)}
& -224.258
& -224.313
& -224.265
& -224.309 \\ \hline\hline
\end{tabular}
\label{tab:O3_wfchar}
\end{table*}

\begin{figure}
    \includegraphics[scale=0.15]{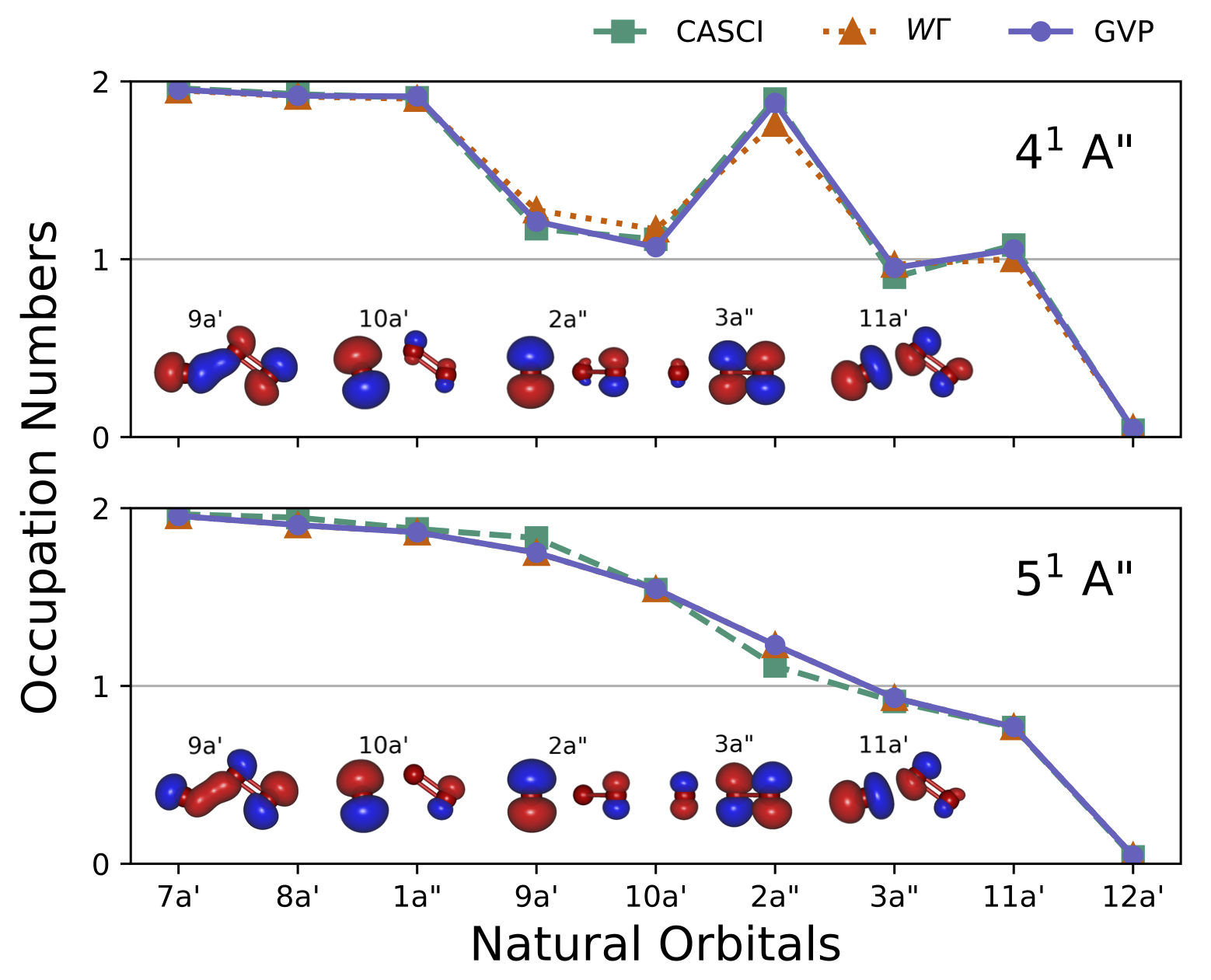}
    \caption{Natural orbital occupation numbers for the $4^{1}A"$ and $5^{1}A"$ excited states of O\textsubscript{3}, calculated from the initial CASCI roots and using the $W\Gamma$ and GVP approaches. The insets show the natural orbitals of each state as calculated by the GVP. For each state, $W\Gamma$ converged the orbital energy gradient to $10^{-4}$ while the GVP converged to $10^{-7}$, leading to small discrepancies in the calculated properties.}
    \label{fig:O3_natorbocc}
\end{figure}

As seen in Figure \ref{fig:O3_natorbocc}, the initial CASCI states (when
swapped in energy ordering) have very similar natural orbital occupation
patterns as the SS-CASSCF energy stationary points,
but a close inspection of the data in Table \ref{tab:O3_wfchar}
suggests that the story is not entirely straightforward.
Indeed, although the GVP optimization starting from the 5th CASCI
root converges tightly and without incident to an energy stationary
point, the final non-orthogonal-CI-style overlaps between this
stationary point and the two CASCI roots (Table \ref{tab:O3_wfchar})
show that a non-trivial remixing has occurred.
The stationary point is still dominated by the CASCI root we started
from (overlap 0.87), but contains a significant amount of the other
root as well (overlap 0.41).

When attempting the GVP optimization starting from the 4th CASCI root,
the story is even less straightforward, with our first attempt at
minimizing the GVP failing to find a stationary point at all.
While this difficulty eventually revealed itself to be an example
of a bad initial wave function guess, this was not obvious until we
had later found the $5^{1}A"$ stationary point and could verify that,
indeed, the CASCI guess was pretty far from the mark.
In practice, it will often be prudent to start from
a better initial guess by using an equal or biased weighting in SA-CASSCF.
Here, however, we intentionally keep this poor initial guess
in order to investigate the efficacy of adding additional
properties to the GVP to help guide the optimization
into the correct basin of convergence.

One property beyond energetics that we can exploit is the fact that
different Hamiltonian eigenstates should be orthogonal to each other.
When using state-specific optimization and an approximate ansatz, this
property will not hold exactly, but should hold approximately.
To help find the $5^{1}A"$ stationary point, we therefore append an
additional component to $\vec{d}$ that (approximately) measures the
overlap between the wave function being optimized and the converged GVP $4^{1}A"$ state.
Our expanded targeting vector in our objective function is now
\begin{align}
    \vec{d} =
    \left \{
    \hspace{1.5mm} \braket{\hat{H}} - \omega,
    \hspace{2mm}
    \frac{\hspace{1.3mm} \vec{b} \cdot \vec{c}\hspace{1.3mm}}
         {\left|\vec{c}\right|}
    \hspace{1.5mm} \right \}
    \label{eq:expanded_d}
\end{align}
in which $\vec{c}$ is the CI vector for the wave function being
optimized and $\vec{b}$ is the normalized CI vector for the converged $4^{1}A"$ stationary point.
The new component is only an approximation to the wave function
overlap, of course, as it does not account for differences in
the shapes of the molecular orbitals in the two wave functions.
However, we do not need it to be exact.
We only need it to be good enough to push the optimization into
the basin of convergence for the $5^{1}A"$ stationary point,
so that when $\mu$ goes to zero in the final stage of GVP
optimization, correct convergence is achieved.

Using the expanded targeting vector from Eq.\ (\ref{eq:expanded_d})
led to a successful GVP optimization in which we again started
from the 4th ${}^{1}A"$ CASCI root, but this time converged
successfully to an energy stationary point for the $5^{1}A"$ state.
As seen from the overlap data in Table \ref{tab:O3_wfchar}, while the primary excitation character is easily assignable to the 4th CASCI root, mathematically this
stationary point is essentially an equal superposition
of the 4th and 5th CASCI roots, revealing that the states
remix strongly during state-specific orbital relaxation and that
the 4th CASCI root really was
a poor initial guess.
Near such a crossing of states, small relaxations of the orbital
shapes can lead to large changes in the CI coefficients.
While the diagonalization procedure of $W\Gamma$ is capable of such changes,
GVP is a local search method and thus finds them challenging without the
help of additional properties.
This motivates more work exploring the abilities of the GVP near energetic
crossings and also in seeding it with equal or biased-weighted SA-CASSCF
starting points that can start us closer to the solution.

In the end, the two energy stationary points that our GVP finds
are made from different mixtures of the 4th and 5th
CASCI roots, although with somewhat relaxed orbitals.
These stationary points are substantially different from
each other but not entirely orthogonal:
their exact NOCI-style overlap with
each other is 0.3, which is not huge but is not zero either.
Thus, although the GVP was successfully able to find SS-CASSCF
stationary points for both states in this difficult case, the fact
that the final stationary points are not as strongly orthogonal
as we might like suggests that the chosen active space could
do with enlargement, or at least that a NOCI
re-diagonalization of these stationary points may be worthwhile.

\subsection{MgO}

\begin{figure}
    \includegraphics[scale=0.135]{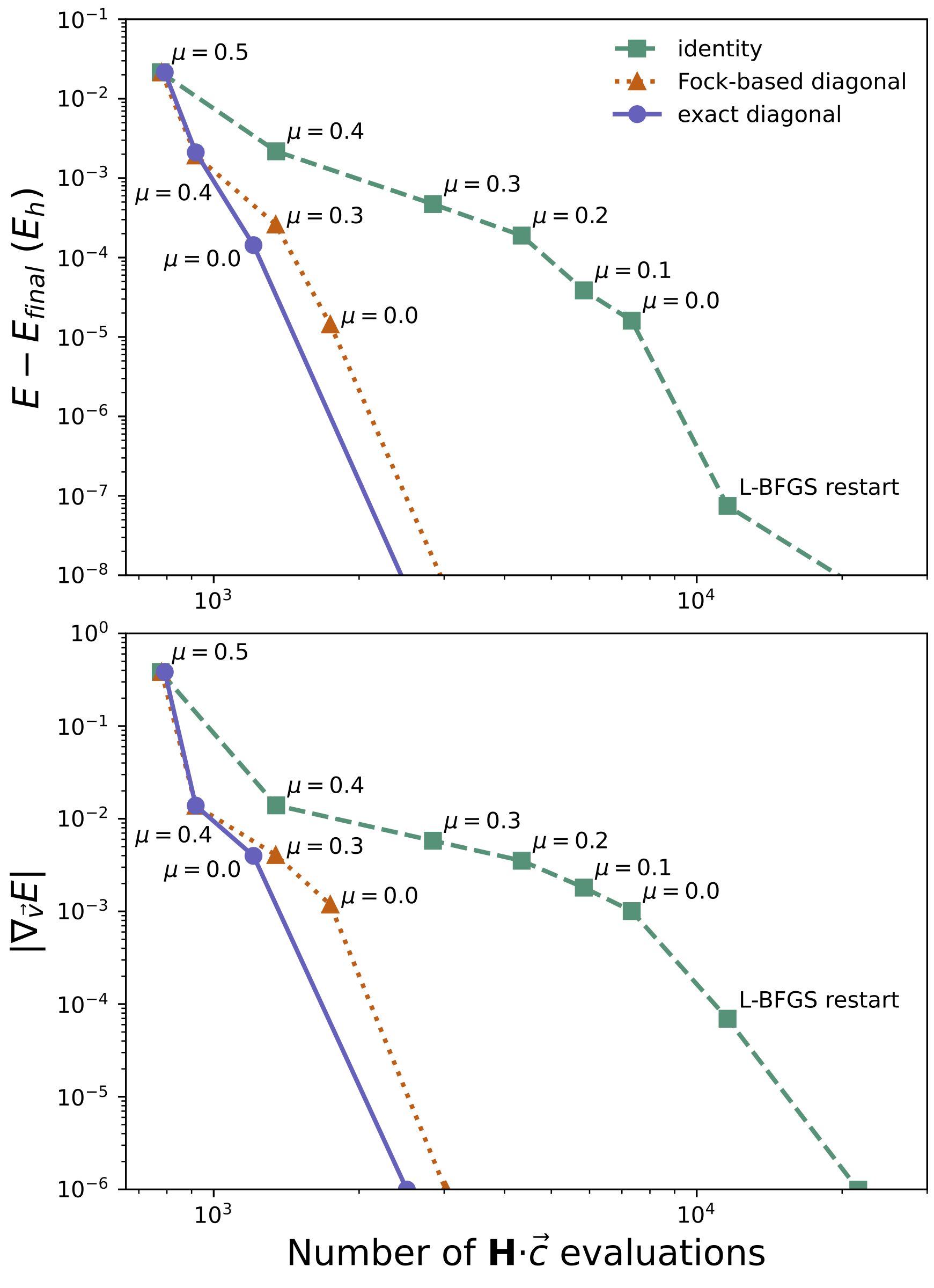}
    \caption{Convergence in terms of energy (top) and energy gradient
    with respect to the variational parameters (bottom) vs the number of Hamiltonian-CI vector contractions 
    for GVP optimizations of the V1 state of MgO.
    Convergence when L-BFGS starts with an approximate Hessian guess built from the exact diagonal (solid purple line) or Fock-based approximate diagonal (dotted orange line) energy Hessian, is compared to convergence when the identity is used instead (green dashed line).
    Starting points for new macro-iterations are labeled. The step down in value of $\mu$ differs between the GVP variations, as determined by the criteria described in Section \ref{section:opt}.
    For all optimizations, the first macro-iteration (not shown) uses the
    identity, $\mu=0.5$, and freezes the CI parameters to provide some
    initial orbital relaxation.}
    \label{fig:MgO_V1convergence}
\end{figure}

\begin{figure}
    \includegraphics[scale=0.14]{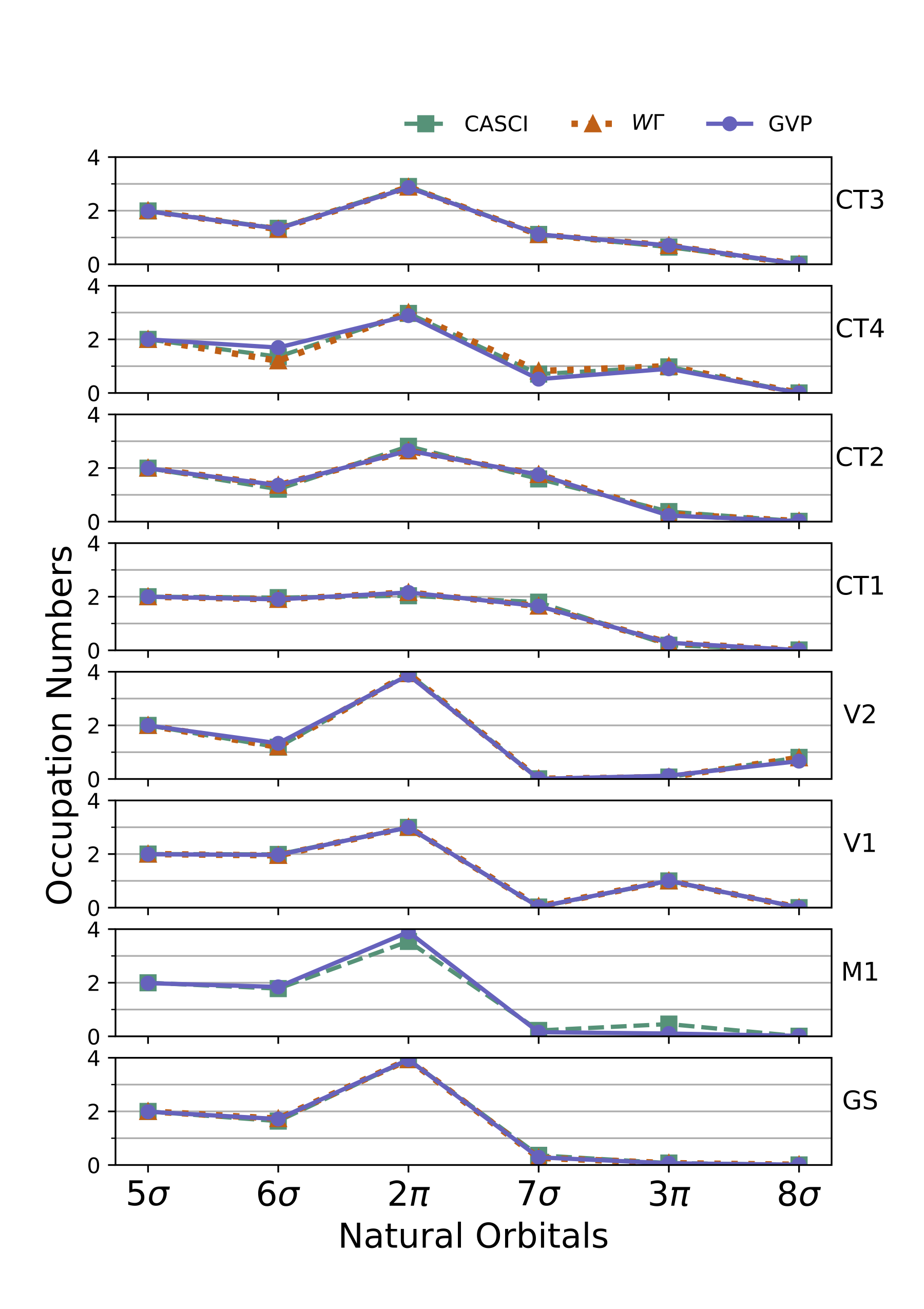}
    \caption{Natural orbital occupation numbers for the first eight ${}^1A_1$
    states in MgO, optimized starting from a CASCI-LDA guess
    with both the $W\Gamma$ and GVP approaches.
    From bottom to top, the states are displayed in ascending order of the
    CASCI-LDA energies, although note that due to orbital relaxation,
    this ordering is not maintained by SS-CASSCF. Note that for both the $2\pi$ and the $3\pi$ labels, there are two symmetry-equivalent spatial orbitals (i.e. $\pi_x$ and $\pi_y$) and we have grouped them such that for these labels the natural orbital occupations range from 0 to 4.}
    \label{fig:MgO_natorbocc}
\end{figure}

\definecolor{Gray}{gray}{0.9} 

\begin{table*}[htbp]\centering
\begin{threeparttable}
\caption{Wavefunction data for ${}^{1}A_{1}$ states in MgO,
listed from top to bottom in ascending order of the CASCI-LDA energies.
Labels (GS, M1, etc) are taken from a previous study. \cite{2step}
The data include the CASCI-LDA dipole moments $\mu$,
wavefunction weight percentages on major components in the LDA orbital basis
(the sum of squared determinant coefficients for all determinants
of the indicated character),
the exact NOCI-style overlaps between the SS-CASSCF stationary points
and the initial CASCI-LDA wavefunctions,
and the predicted excitation energies.
} 
\small
\begin{tabular}{c c| c r c l | c c c | c c | c c c }
\hline\hline
& & & & & & \multicolumn{3}{c|}{\underline{Wavefunction Weight \%}} & \multicolumn{2}{c|}{\underline{Overlap}} & \multicolumn{3}{c}{\underline{Excitation E (eV)}} \\
State & Label & $\mu$ (D) & \multicolumn{3}{c|}{Primary Excitations} & CASCI & $W\Gamma$ & GVP & $W\Gamma$ & GVP & CASCI & $W\Gamma$ & GVP \\ \hline
\multirow{2}{*}{$1^1A_1$} & \multirow{2}{*}{GS} & \multirow{2}{*}{-3.95} & \multicolumn{3}{c|}{Aufbau} &  76.5 & 81.9 & 81.9 & \multirow{2}{*}{0.95} & \multirow{2}{*}{0.95} & \multirow{2}{*}{0} & \multirow{2}{*}{0} & \multirow{2}{*}{0} \\ 
 & & & 6$\sigma^2$ & $\rightarrow$ & 7$\sigma^2$ &  12.1 & 10.9 & 10.9 &  & & & &\\
 & & & & & & & & & & & & &\\
 \multirow{3}{*}{$2^1A_1$} & \multirow{3}{*}{M1} & \multirow{3}{*}{-5.39} & 6$\sigma$ & $\rightarrow$ & 7$\sigma$ & 41.8 & -- & 56.1 & \multirow{3}{*}{--} & \multirow{3}{*}{0.80} & \multirow{3}{*}{2.48} & \multirow{3}{*}{--} & \multirow{3}{*}{3.11}\\
 & & & 2$\pi$ & $\rightarrow$ & 3$\pi$ & 25.2 & -- & 1.0 & &  & & &\\ 
 & & & 6$\sigma^2$ & $\rightarrow$ & 7$\sigma^2$ &  15.1 & -- & 38.4 &  & & & &\\ 
 & & & & & & & & & & & & &\\
\multirow{2}{*}{$3^1A_1$} & \multirow{2}{*}{V1} & \multirow{2}{*}{-4.88} & 2$\pi$ & $\rightarrow$ & 3$\pi$ & 68.4 & 72.5 & 72.2 & \multirow{2}{*}{0.98} & \multirow{2}{*}{0.98} & \multirow{2}{*}{3.70} & \multirow{2}{*}{4.88} & \multirow{2}{*}{4.88}\\ 
 & & & 6$\sigma$, 2$\pi$ & $\rightarrow$ & 7$\sigma$, 3$\pi$ & 22.3 & 19.8 & 20.0 & & & & &\\
 & & & & & & & & & & & & &\\
\multirow{7}{*}{$4^1A_1$} & \multirow{7}{*}{V2} & \multirow{7}{*}{-5.93} & 6$\sigma$ & $\rightarrow$ & 8$\sigma$ & 70.5 & 44.1 & 59.9 & \multirow{7}{*}{0.35} & \multirow{7}{*}{0.96} & \multirow{7}{*}{6.46} & \multirow{7}{*}{6.60} & \multirow{7}{*}{8.25}\\ 
 & & & 6$\sigma^2$ & $\rightarrow$ & 7$\sigma$, 8$\sigma$ & 14.8 & 22.8 & 15.0 & & & & &\\ 
 & & & 6$\sigma$, 2$\pi$ & $\rightarrow$ & 3$\pi$, 8$\sigma$ & 5.3 & 3.8 & 4.4 & & & & &\\
 & & & 2$\pi$ & $\rightarrow$ & 3$\pi$ & 3.9 & 0.7 & 9.1 & & & & &\\
 & & & 6$\sigma$, 2$\pi$ & $\rightarrow$ & 7$\sigma$, 3$\pi$ & 2.1 & 0.6 & 5.2 & & & & &\\
 & & & \multicolumn{3}{c|}{Aufbau} & 0.4 & 17.7 & 1.0 & & & & &\\
 & & & 6$\sigma^2$ & $\rightarrow$ & 7$\sigma^2$ & 0.4 & 6.0 & 0.5 & & & & &\\
 & & & & & & & & & & & & &\\
\multirow{3}{*}{$5^1A_1$} & \multirow{3}{*}{CT1} & \multirow{3}{*}{3.84} & 2$\pi^2$ & $\rightarrow$ & 7$\sigma^2$ & 62.8 & 60.7 & 60.5 & \multirow{3}{*}{0.92} & \multirow{3}{*}{0.92} & \multirow{3}{*}{7.15} & \multirow{3}{*}{6.57} & \multirow{3}{*}{6.57}\\ 
 & & & 2$\pi^2$ & $\rightarrow$ & 7$\sigma$, 8$\sigma$ & 13.3 & 7.4 & 7.5 &  & & & &\\ 
 & & & 2$\pi^3$ & $\rightarrow$ & 7$\sigma^2$, 3$\pi$ & 8.7 & 7.8 & 7.8 &  & & & &\\ 
 & & & & & & & & & & & & &\\
\multirow{3}{*}{$6^1A_1$} & \multirow{3}{*}{CT2} & \multirow{3}{*}{3.93} & 2$\pi^2$ & $\rightarrow$ & 7$\sigma^2$ & 30.2 & 44.0 & 44.1 & \multirow{3}{*}{0.91} & \multirow{3}{*}{0.91} & \multirow{3}{*}{7.62} & \multirow{3}{*}{7.30} & \multirow{3}{*}{7.30}\\ 
 & & & 6$\sigma^2$ & $\rightarrow$ & 7$\sigma^2$ & 16.9 & 14.7 & 14.7 & & & & &\\ 
 & & & 6$\sigma$, 2$\pi$ & $\rightarrow$ & 7$\sigma$, 3$\pi$ & 13.9 & 8.0 & 8.0 & & & & &\\ 
 & & & & & & & & & & & & &\\
    \multirow{2}{*}{$7^1A_1$} & \multirow{2}{*}{CT4} & \multirow{2}{*}{2.33} & 6$\sigma$, 2$\pi$ & $\rightarrow$ & 7$\sigma$, 3$\pi$ & 47.1 & 70.7 & 52.6 & \multirow{2}{*}{0.30} & \multirow{2}{*}{0.90} & \multirow{2}{*}{8.07} & \multirow{2}{*}{11.65} & \multirow{2}{*}{8.69} \\ 
 & & & 6$\sigma^2$, 2$\pi$ & $\rightarrow$ & 7$\sigma^2$, 3$\pi$ & 27.0 & 13.1 & 24.5 & & & & &\\ 
 & & & & & & & & & & & & &\\
\multirow{6}{*}{$8^1A_1$} & \multirow{6}{*}{CT3} & \multirow{6}{*}{3.66} & 2$\pi$ & $\rightarrow$ & 3$\pi$ & 19.0 & 16.0 & 7.9 & \multirow{6}{*}{0.91} & \multirow{6}{*}{0.88} & \multirow{6}{*}{8.16} & \multirow{6}{*}{8.39} & \multirow{6}{*}{8.54} \\ 
 & & & 6$\sigma$, 2$\pi$ & $\rightarrow$ & 7$\sigma$, 3$\pi$ & 17.4 & 23.8 & 31.6 & & & & &\\ 
 & & & 2$\pi^2$ & $\rightarrow$ & 7$\sigma^2$ & 16.6 & 12.8 & 17.3 & & & & &\\ 
 & & & 6$\sigma$ & $\rightarrow$ & 7$\sigma$ & 10.0 & 4.0 & 1.6 & & & & &\\
 & & & 6$\sigma^2$, 2$\pi$ & $\rightarrow$ & 7$\sigma^2$, 3$\pi$ & 8.6 & 10.1 & 9.9 & & & & &\\ 
 & & & 2$\pi^2$ & $\rightarrow$ & 3$\pi^2$ & 5.9 & 8.0 & 6.0 & & & & &\\  \hline\hline
\end{tabular}
\label{tab:MgO_wfchar}
\end{threeparttable}

\end{table*}

\begin{table}[htbp]
    \caption{Two representative attempts at achieving SS-CASSCF convergence
    in MgO's M1 state by SA-CASSCF with shifting weights via
    Molpro version 2019.2 with default SA-CASSCF optimizer settings
    (aside from the use of biased SA weights).
    Each attempt starts with an equal-weight SA-CASSCF (seeded with LDA orbitals)
    and then, for each additional row in the table, uses the previous SA-CASSCF's
    result as the guess for a new calculation with more biased weights.
    A 4-state SA was used to simplify the problem by avoiding the states with CT
    character, but even with this simplification we were not able to get closer than
    having about 90\% of the weight on the target state before root flipping prevented
    SA-CASSCF from converging.
    The converged SS-CASSCF energy for M1 found by GVP is -274.403367 E$_h$.
    }
    \small
    \centering
    \begin{tabular}{c c c c c}
    \hline
    \hline
    \multicolumn{5}{c}{Attempt 1} \\[0.2mm]
    Energy (E$_h$) $\quad$ & Weight 0 $\quad$ &  Weight 1   $\quad$ &  Weight 2  $\quad$ &  Weight 3  \\
    \hline
   -274.371506  $\quad$ &  0.250  $\quad$ &  0.250  $\quad$ &  0.250 $\quad$ &  0.250 \\
   -274.376089  $\quad$ &  0.200  $\quad$ &  0.400  $\quad$ &  0.200 $\quad$ &  0.200 \\
   -274.384705  $\quad$ &  0.100  $\quad$ &  0.700  $\quad$ &  0.100 $\quad$ &  0.100 \\
   -274.396991  $\quad$ &  0.050  $\quad$ &  0.900  $\quad$ &  0.050 $\quad$ &  0.000 \\
   no convergence     $\quad$ &  0.025  $\quad$ &  0.950  $\quad$ &  0.025 $\quad$ &  0.000 \\[1mm]
    \multicolumn{5}{c}{Attempt 2} \\[0.2mm]
    Energy (E$_h$) $\quad$ & Weight 0 $\quad$ &  Weight 1   $\quad$ &  Weight 2  $\quad$ &  Weight 3  \\
    \hline
   -274.371506    $\quad$ & 0.250    $\quad$ & 0.250    $\quad$ & 0.250    $\quad$ & 0.250 \\
   -274.378506    $\quad$ & 0.300    $\quad$ & 0.400    $\quad$ & 0.200    $\quad$ & 0.100 \\
   -274.390694    $\quad$ & 0.300    $\quad$ & 0.600    $\quad$ & 0.050    $\quad$ & 0.050 \\
   -274.390525    $\quad$ & 0.400    $\quad$ & 0.600    $\quad$ & 0.000    $\quad$ & 0.000 \\
   -274.392357    $\quad$ & 0.300    $\quad$ & 0.700    $\quad$ & 0.000    $\quad$ & 0.000 \\
   -274.394417    $\quad$ & 0.200    $\quad$ & 0.800    $\quad$ & 0.000    $\quad$ & 0.000 \\
   -274.397643    $\quad$ & 0.100    $\quad$ & 0.900    $\quad$ & 0.000    $\quad$ & 0.000 \\
    no convergence      $\quad$ & 0.050    $\quad$ & 0.950    $\quad$ & 0.000    $\quad$ & 0.000 \\
    \hline
    \hline
    \end{tabular}
    \label{tab:sa_mgo_attempts}
\end{table}

\begin{figure*}
    \includegraphics[scale=0.15]{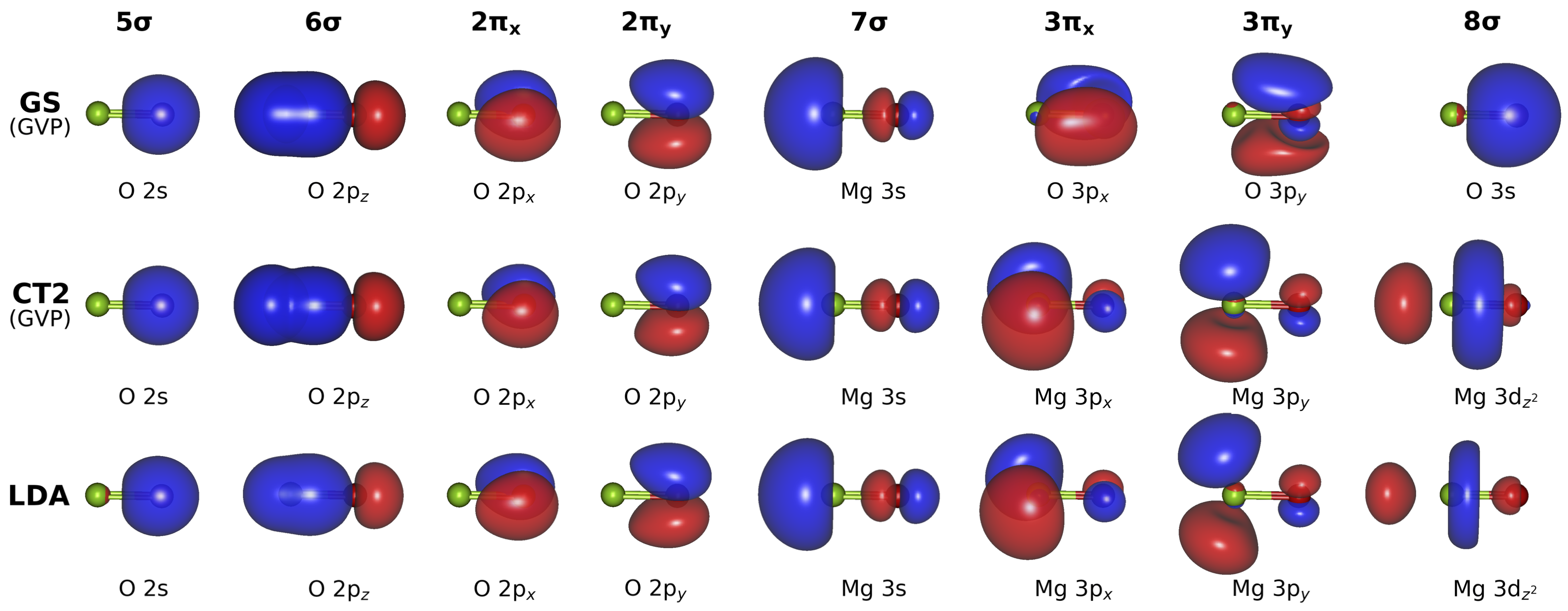}
    \caption{The MgO active orbitals in
    the LDA guess (bottom row) and the SS-CASSCF stationary points
    for CT2 (middle row) and the ground state (top row).
    Each image has the Mg atom at left in green and text
    indicating the orbital's primary character.
    }
    \label{fig:MgO_LDAorbs}
\end{figure*}

As our third and final example, we use the GVP to find SS-CASSCF energy
stationary points corresponding to each of the eight lowest ${}^1$A$_1$
CASCI roots in MgO at a bond length of $1.8$ \text{\AA} and
with an (8o, 8e) active space.
The excited states in MgO present a challenging array of multi-reference
and charge transfer character, \cite{MgO-1,MgO-2,MgO-3}
as can be seen from an inspection of
Table \ref{tab:MgO_wfchar} and Figures \ref{fig:MgO_natorbocc} and \ref{fig:MgO_LDAorbs}.
Some states exhibit both behaviors at once, such as the CT2 state, which is a
doubly-excited, double-charge-transfer state in which the most prominent
electron configuration accounts for less than half the wave function.
SS-CASSCF is an especially appropriate theory in this setting,
being able to deal with both the strong post-CT orbital relaxation
and the multi-reference character that so often comes along with
double excitations.
Previous work with state-averaged CASSCF has investigated the lowest
excited state in MgO, \cite{MgO-SA}
and in principle dynamic weighting \cite{dynamicSA}
may be able to help in making predictions about the others,
but the mix of neutral and ionic character in these states
makes standard state averaging hard to recommend, and if one
wishes to take dynamic weighting to its limit, one is really
asking for SS-CASSCF.
However, even when SS-CASSCF is the goal, the method of optimization
matters a great deal, with a previous study showing that simple root
selection fails to converge to the initially targeted state in
state-specific optimizations of all seven of the lowest
${}^1$A$_1$ excited states. \cite{2step}
Similarly, we find that a shifting-weight
SA-CASSCF approach struggles with root flipping in some of these states,
as shown in Table \ref{tab:sa_mgo_attempts}.
Using a careful analysis based on NOCI overlaps,
we find that, while the W$\Gamma$ optimization method is more
effective, it still fails to locate an appropriate stationary point
for three of these seven excited states.
By adding the GVP approach to our toolbox, however, we are able to find
good energy stationary points for the ground state and all seven excited states.

Before getting into the state-by state details, let us first emphasize
the value of supplying L-BFGS with our approximate diagonal form for
the initial objective function Hessian as opposed to the identity matrix.
For this comparison, as for all the optimizations in this section,
our starting point is a particular root from a CASCI calculation
carried out in the LDA orbital basis (denoted as CASCI-LDA),
with the active space chosen
as the lowest four LDA orbitals of $\sigma$ character plus the lowest
four of $\pi$ character, as seen in Figure \ref{fig:MgO_LDAorbs}.
These active orbitals can be roughly characterized as the
O 2s and 2p and the Mg 3s, off-axis 3p, and 3d$_{z^2}$ orbitals.
The Mg 1s, 2s, and 2p and the O 1s orbitals are held closed but not frozen.
As seen in Figure \ref{fig:MgO_V1convergence}, employing
either version of our diagonal Hessian
approximation speeds up the optimization convergence for the V1 state
by more than an order of magnitude relative to using the identity matrix.
Similar speed ups were observed for other states as well.
There is still room for improvement, however, and so in future
it will be interesting to investigate combinations of GVP-based
L-BFGS with more standard tools like Davidson CI steps
and more traditional orbital optimizations.

Turning now to stability, we find that, with this new GVP optimization
method in hand, we can now locate stationary points for all eight of
the lowest ${}^1$A$_1$ states, as shown in Table \ref{tab:MgO_wfchar}.
The ground state is the simplest, and indeed all optimization methods
-- including GVP, W$\Gamma$, SRS, and the default PySCF ground state
CASSCF solver -- come to the same stationary point.
The lowest excited state (M1) is a more significant case, as no previous
method has to our knowledge been able to locate the full (orbital + CI)
energy stationary point for this state.
Despite its careful root tracking approach, W$\Gamma$ collapses to
the ground state when trying to target the M1 state starting from
the corresponding CASCI-LDA root.
In contrast, GVP has no trouble with this state, finding a stationary
point that, based on its NOCI overlap with the starting CASCI-LDA root,
clearly corresponds to the excited state being sought.
Turning to the V1 and CT2 states, both GVP and W$\Gamma$ work well,
arriving at the same stationary points that, again, have large overlaps
with the CASCI-LDA excited states used to initiate the optimizations
and define which excited state we are after.
The V2 and CT4 states both represent failures for the W$\Gamma$ approach,
however, which was not obvious in the previous study \cite{2step}
as a natural orbital occupation analysis (Figure \ref{fig:MgO_natorbocc})
makes it appear that the stationary points arrived at are a match for
the states being sought.
However, NOCI overlaps, which we have now evaluated and which are a more
direct measure of wave function similarity, show that in both V2 and CT4,
W$\Gamma$ converges to a stationary point that is of a very different
character than the excited state in question.
GVP, on the other hand, finds stationary points for these states
that have large overlaps with the starting CASCI-LDA roots and
so clearly match the states being sought.
In CT1, we have our one example in MgO in which the simplest
use of the GVP (energy targeting only) fails to find a
stationary point, the optimization getting stuck 
at an energy gradient norm of roughly $10^{-4}$.
However, W$\Gamma$ works in this case, and GVP can be improved either by
expanding the vector $\vec{d}$, as we did in the upper ozone state,
or by improving the initial guess, which is the approach we take here.
If we supply slightly better orbitals by taking them from the output
of the second macro-iteration of W$\Gamma$ (but still using the
CASCI-LDA CI vector guess so as not to give GVP too much help)
we find that the GVP optimization is able to converge to the
same stationary point as found by W$\Gamma$.
The final state we are looking at, CT3, is an even more interesting
case, in which W$\Gamma$ and GVP find two different stationary points,
both of which have strong overlap with the sought after state.
The difference between these stationary points is in the
8$\sigma$ orbital, which in the GVP stationary point has O 3s character
but in the W$\Gamma$ stationary point has Mg 3d$_{z^2}$ character.
Given their large overlaps with the initial CASCI-LDA root and
their large overlap of 0.93 with
each other, they both appear to be approximations of the same Hamiltonian
eigenstate and thus a good example of how nonlinear wave function forms
can have more stationary points than there are physical eigenstates.
Rather than try to choose between them, we see this as a case that indicates
the active space is, at least for this state, at least one orbital too small.

As in other types of CASSCF, multiple solutions can exist when
the highest energy active orbitals are only slightly occupied and
it is possible to get similarly good wave functions when swapping one
or more of them with low-lying virtual orbitals.
This issue can cause multiple nearby minima in both ground state and SA-CASSCF, although it is entirely case by case whether
swaps between the least occupied active orbitals and the lowest
virtual orbitals move the optimization between different
local minima or simply move it around within the
same basin of convergence surrounding a single minimum.
Our results for CT3 provide evidence that something like
the multiple-minima issue can occur for excited states in SS-CASSCF,
with two very similar stationary points differing by a
swap between low-lying virtuals and high-lying active orbitals.
In the case of CT3, 
one might prefer the 3d$_{z^2}$ stationary point on the basis
that it contains only valence orbitals in its active space, but
applying such logic in general is not straightforward.
Indeed, all optimization methods we have tried (including the
default implementation in PySCF) agree that, after state-specific
optimization, the ground state active space displayed in Figure \ref{fig:MgO_LDAorbs} contains orbitals
with O 3s, 3p$_x$, and 3p$_y$ character, having swapped them in for
the LDA Mg 3p$_x$, 3p$_y$, and 3d$_{z^2}$ valence orbitals that were
present in the the initial guess.
What is essentially going on here is that, if only a subset of
the active orbitals need to have significant occupation in order
to capture the strong correlation effects in a given state,
then, for that state, the choice for the remaining active
orbitals that will give the lowest energy is whichever ones
provide the best ability to capture some weak correlation,
and there is no particular reason that these will be valence orbitals.
In the ground state, it makes some sense for the O 3-shell
orbitals to be more effective for this purpose than the unoccupied
Mg valence orbitals, as the ground state concentrates the electrons
on the O atom, putting a premium on orbitals that can help describe
weak correlation effects in its vicinity.
Another well-known example of this issue, although not in play here,
is the double d-shell effect,\cite{dshell-1,dshell-2} where it is often wise to include
non-valence d orbitals in the active space for transition metal
compounds ahead of some orbitals that are formally valence orbitals.
As in ground states or state averaging cases with multiple minima,
the best approach to removing the ambiguity between CT3's two
stationary points is probably to expand the active space.
By doing so, the orbitals that are competing for inclusion in
the active space and leading to multiple stationary points can
all be included, at which point we expect the two stationary
points would merge into one.
From an optimization perspective, this would amount to the two
minima on the $|\nabla_{\vec{v}}E|^2$ surface joining into
a single minimum with a single basin of convergence.
Certainly this must happen in the limit that the active space
expands CASSCF into FCI, but we suspect that in this case it will
happen immediately upon allowing both the O 3s and Mg 3d$_{z^2}$
orbitals to be in the active space simultaneously.

\section{Conclusion}

We have shown that excited-state-specific optimization of the CASSCF ansatz
via the minimization of a generalized variational principle allows
the desired excited state stationary points to be located and tightly converged in multiple challenging scenarios.
The GVP consists of the square norm of the energy gradient along with a
steering term that allows approximately known properties of the desired state
to guide the optimization to its energy stationary point.
The form permits a very broad variety of properties to be employed, and in
this study we have used estimates for the energy and, in one particularly
challenging case, rough orthogonality against another state for this purpose.
By achieving state-specific optimization with the GVP, situations where this approach could be especially helpful include cases where state-averaging is frustrated by root flipping, high-lying states where it is not practical to resolve all lower-lying states, avoided crossings, and states displaying both strongly correlated character and strong orbital relaxations, as in some core, charge transfer and doubly excited states.

In our results, we find that the GVP approach is capable of converging to the
correct stationary point in excited states of LiH, ozone, and MgO in which
root flipping is present.
Its tighter convergence than uncoupled two-step methods produces energies in
LiH that are significantly closer to FCI, and its root-targeting capabilities
allow it to match the efficacy of the recently developed W$\Gamma$ method in a
nearly degenerate pair of states in ozone.
In MgO, it was not previously possible to find the correct stationary points
for three excited singlet states in the symmetric representation of the computational
point group.
With the addition of the GVP approach, all three of these missing stationary points
have been found.

Looking forward, there are a number of promising directions worth pursuing.
First, this study limited itself to using quasi-Newton optimization of the GVP
objective function, which is illuminating but almost certainly not the most
efficient approach given the historical dominance of the Davidson algorithm
when dealing with CI coefficients.
Methods that combine the flexibility and reliability of GVP minimization with the
efficiency of Krylov subspace eigensolvers are thus a priority for future
method development.
If sticking with a quasi-Newton approach, directions to consider for improving optimization efficiency include correcting the L-BFGS gradient history when shifting the orbital reference throughout the optimization, as well as delving into approximate initial Hessians that retain more of the CI-orbital coupling.
Second, CASSCF energetics are rarely quantitative due to a lack of treatment
of weak correlation effects.
With the GVP approach able to provide excited state stationary points in a wider
range of cases than was previously possible, it will be interesting to perform
more extensive tests on what benefits this can offer to post-CASSCF weak
correlation methods.
Whatever these directions uncover, it is becoming increasingly clear that it
is possible and often desirable to achieve fully excited-state-specific
quantum chemistry in a wide variety of single-reference and
multi-reference methods.

\section{Acknowledgements}

This work was supported by the National Science Foundation's
CAREER program under Award Number 1848012. Calculations were performed 
using the Berkeley Research Computing Savio cluster and the Lawrence 
Berkeley National Lab Lawrencium cluster. R.H. acknowledges that this 
material is based upon work supported by the National Science Foundation 
Graduate Research Fellowship Program under Grant No. DGE 1752814 and 
DGE 2146752. Any opinions, findings, and conclusions or recommendations 
expressed in this material are those of the authors and do not necessarily 
reflect the views of the National Science Foundation.


\section{Supporting Information}
\renewcommand{\thesection}{S\arabic{section}}
\renewcommand{\theequation}{S\arabic{equation}}
\renewcommand{\thefigure}{S\arabic{figure}}
\renewcommand{\thetable}{S\arabic{table}}
\setcounter{section}{0}
\setcounter{figure}{0}
\setcounter{equation}{0}
\setcounter{table}{0}
\section{Orbital Energy Derivatives}
For the orbital block of the energy derivatives, we define
$E_{pq}^- =
\left( \hat{a}_{p}^{\dag} \hat{a}_{q} - \hat{a}_{q}^{\dag} \hat{a}_{p} \right)$
and $P_{pq,rs}$ as a permutation operator giving us the following expressions for the orbital energy gradient and Hessian.
\begin{align}
    \frac{\partial E}{\partial X_{pq}} &=
    \bra{\Psi} \left[ E_{pq}^-, \hat{H} \right] \ket{\Psi}
    \label{eq:oGrad} \\
    \frac{\partial^2 E}{\partial X_{pq} \partial X_{rs}} &=
    \frac{1}{2} (1+P_{pq,rs}) \bra{\Psi} \left[ E_{pq}^- , \left[ E_{rs}^- , \hat{H} \right] \right] \ket{\Psi}
    \label{eq:ooHess_SI}
\end{align}
Core orbitals are indexed using $i,j,k$, active orbitals with $t,u,v,w$, and virtual with $a,b,c$ where $p,q,r,s$ are used for general orbitals. For simplicity, we define several Fock-type matrices:
\begin{align}
    & F_{pq}^{core} = h_{pq} + \sum_{k}^{N_{core}} \left[ 2 (pq|kk) - (pk|kq)\right] \\
    & F_{pq}^{act} = \sum_{uv}^{N_{act}} \gamma_{uv} \left[ 2 (pq|uv) - (pu|vq)\right] \\
    & F_{pq}^{occ} = h_{pq} + \sum_{r}^{N_{occ}} \left[ 2 (pq|rr) - (pr|rq)\right]
\end{align}
where the one and two-electron spin-summed reduced density matrices are defined as
\begin{align}
    \gamma_{pq} &= \sum_{IJ} c_I c_J \bra{\phi_I} \hat{a}_{p}^{\dag} \hat{a}_{q} \ket{\phi_J} \\
    &= \sum_{IJ} c_I c_J \bra{\phi_I} \left( \hat{a}_{p_{\alpha}}^{\dag} \hat{a}_{q_{\alpha}} + \hat{a}_{p_{\beta}}^{\dag} \hat{a}_{q_{\beta}}\right) \ket{\phi_J} \notag \\
    \Gamma_{pqrs} &= \sum_{IJ} c_I c_J \bra{\phi_I} \hat{a}_{p}^{\dag} \hat{a}_{r}^{\dag} \hat{a}_{s} \hat{a}_{q} \ket{\phi_J} \\
    &= \sum_{IJ} c_I c_J \bra{\phi_I} \Big( \hat{a}_{p_{\alpha}}^{\dag} \hat{a}_{r_{\alpha}}^{\dag} \hat{a}_{s_{\alpha}} \hat{a}_{q_{\alpha}} + \hat{a}_{p_{\alpha}}^{\dag} \hat{a}_{r_{\beta}}^{\dag} \hat{a}_{s_{\beta}} \hat{a}_{q_{\beta}}\\
    &\qquad + \hat{a}_{p_{\alpha}}^{\dag} \hat{a}_{r_{\beta}}^{\dag} \hat{a}_{s_{\beta}} \hat{a}_{q_{\alpha}} + \hat{a}_{p_{\beta}}^{\dag} \hat{a}_{r_{\alpha}}^{\dag} \hat{a}_{s_{\alpha}} \hat{a}_{q_{\beta}} \Big) \ket{\phi_J}. \notag
\end{align}

\subsection{Orbital Energy Gradient}
Using the index definitions and Fock-type matrices defined in the previous section, the exact expressions for the core-virtual, active-virtual, and core-active blocks of the orbital energy gradient in Eq.\ (\ref{eq:oGrad}) evaluated at $X=0$ are as follows.
\begin{align}
    \frac{\partial E}{\partial X_{ia}} &= 4F_{ai}^{core} + 2F_{ai}^{act}
    \label{eq:oGrad_1}\\
    \frac{\partial E}{\partial X_{ta}} &= 2\sum_u^{N_{act}}\gamma_{tu}F_{au}^{core} + 2\sum_{uvw}^{N_{act}} \Gamma_{tuvw}(au|vw)
    \label{eq:oGrad_2}\\
    \frac{\partial E}{\partial X_{it}} &= 4F_{ti}^{core} + 2F_{ti}^{act} \\
    &\qquad - 2\sum_{u}^{N_{act}} \gamma_{tu}F_{iu}^{core} - 2\sum_{uvw}^{N_{act}} \Gamma_{tuvw}(iu|vw)
    \label{eq:oGrad_3}
\end{align}

\subsection{Approximate Orbital Energy Hessian}
\subsubsection{Exact Diagonal}
Taking only the diagonal elements of the energy Hessian, Eq. \eqref{eq:ooHess_SI} simplifies to
\begin{align}
    \frac{\partial^2 E}{\partial X_{pq}^2} =
    \bra{\Psi} \left[ E_{pq}^- , \left[ E_{pq}^- , \hat{H} \right] \right] \ket{\Psi}.
    \label{eq:ooHess_diag}
\end{align}
The following are exact expressions for the core-virtual, active-virtual, and core-active blocks of the diagonal orbital energy Hessian in Eq.\ (\ref{eq:ooHess_diag}) evaluated at $X=0$:
\begin{align}
    \frac{\partial^2 E}{\partial X_{ia}^2} = \hspace{1mm}&4F_{aa}^{core} + 2F_{aa}^{act} - 4F_{ii}^{core} - 2F_{ii}^{act} \notag \\
    &- 4(aa|ii) + 12(ai|ai) \\
    \frac{\partial^2 E}{\partial X_{ta}^2} = \hspace{1mm}&2\gamma_{tt}F_{aa}^{core} - 2 \sum_u^{N_{act}}\gamma_{tu}F_{tu}^{core} - 2\sum_{uvw}^{N_{act}} \Gamma_{tuvw}(tu|vw) \notag\\
    &+ 2\sum_{uv}^{N_{act}} \left[ \Gamma_{tutv}(au|av) + \Gamma_{tvut}(au|av) + \Gamma_{ttvu}(aa|vu) \right] \\
    \frac{\partial^2 E}{\partial X_{it}^2} = \hspace{1mm}&4F_{tt}^{core} + 2F_{tt}^{act} - 4F_{ii}^{core} - 2F_{ii}^{act} \notag \\
    &+ 2\gamma_{tt}F_{ii}^{core} - 2\sum_{u}^{N_{act}} \gamma_{tu}F_{tu}^{core} - 2\sum_{uvw}^{N_{act}} \Gamma_{tuvw}(tu|vw) \notag \\
    &+ 2\sum_{uv}^{N_{act}} \left[\Gamma_{tutv}(ui|iv) + \Gamma_{tvut}(ui|iv) + \Gamma_{ttuv}(uv|ii)\right] \notag \\
    &+ 4\sum_{u}^{N_{act}} \left[3(ui|ui) - (uu|ii) - 3\gamma_{tu}(ui|ti) + \gamma_{tu}(tu|ii) \right].
\end{align}

\subsubsection{Fock-based Approximate Diagonal}
Adding an additional layer of approximation, we go dropping the off-diagonal terms of the energy Hessian and approximate the Hamiltonian inside the commutators with the one-electron Fock operator, giving us the following approximation to Eq.\ (\ref{eq:ooHess_diag}).
\begin{align}
    \frac{\partial^2 E}{\partial X_{pq}^2} \approx \bra{\Psi} \left[ E_{pq}^- , \left[ E_{pq}^- , \hat{F} \right] \right] \ket{\Psi}
    \label{eq:ooHess_diag_approx}
\end{align}
Building the Fock operator from our CASSCF wave function's one-body density matrix and the effective one-electron integrals:
\begin{align}
    \hat{F} = \sum_{pq} \left( h_{pq} + \sum_{r} \left[ 2 (pq|rr) - (pr|rq) \right] \right) \hat{a}_{p}^{\dag} \hat{a}_{q}.  
\end{align}
With this approximation we arrive at the approximate expressions for core-virtual, active-virtual, and core-active blocks of the diagonal orbital energy Hessian in Eq.\ (\ref{eq:ooHess_diag_approx}) evaluated at $X=0$:
\begin{align}
    \frac{\partial^2 E}{\partial X_{ia}^2} &\approx 2F_{aa}^{occ} - 2F_{ii}^{occ} \\
    \frac{\partial^2 E}{\partial X_{ta}^2} &\approx 2F_{aa}^{occ} \gamma_{tt} - 2 \sum_u^{N_{act}}F_{tu}^{occ}\gamma_{tu} \\
    \frac{\partial^2 E}{\partial X_{it}^2} &\approx 2F_{ii}^{occ} \gamma_{tt} + 2F_{tt}^{occ} - 2F_{ii}^{occ} - 2 \sum_u^{N_{act}}F_{tu}^{occ}\gamma_{tu}.
\end{align}

\section{Additional Data}
${}^{}$
\begin{table}[htbp]
\caption{Energies (E$_h$) of the first excited state $A^1\Sigma^+$ of LiH at various bond lengths.}
\small
\begin{tabular}{c c c c }
\hline\hline
R (\AA) & FCI & $W\Gamma$ & GVP \\ \hline
1.2 & $\quad$-7.8421784 & $\quad$-7.8369774 & $\quad$-7.8379204 \\
1.4 & $\quad$-7.8718929 & $\quad$-7.8685677 & $\quad$-7.8689355 \\
1.6 & $\quad$-7.8873115 & $\quad$-7.8843640 & $\quad$-7.8844385 \\
1.8 & $\quad$-7.8950433 & $\quad$-7.8921683 & $\quad$-7.8930879 \\
2.0 & $\quad$-7.8987095 & $\quad$-7.8958730 & $\quad$-7.8968039 \\
2.2 & $\quad$-7.9002698 & $\quad$-7.8973900 & $\quad$-7.8983689 \\
2.4 & $\quad$-7.9007174 & $\quad$-7.8978058 & $\quad$-7.8982932 \\
2.6 & $\quad$-7.9005042 & $\quad$-7.8975273 & $\quad$-7.8979879 \\
2.8 & $\quad$-7.8997797 & $\quad$-7.8966386 & $\quad$-7.8971273 \\
3.0 & $\quad$-7.8985339 & $\quad$-7.8953840 & $\quad$-7.8957249 \\ 
3.4 & $\quad$-7.8931780 & $\quad$-7.8908310 & $\quad$-7.8907296 \\
3.8 & $\quad$-7.8879230 & $\quad$-7.8847120 & $\quad$-7.8846122 \\
4.2 & $\quad$-7.8809573 & $\quad$-7.8783253 & $\quad$-7.8782487 \\
\hline\hline
\end{tabular}
\label{tab:LiH_energies}

\end{table}

\begin{figure}[ht]
    \includegraphics[scale=0.13]{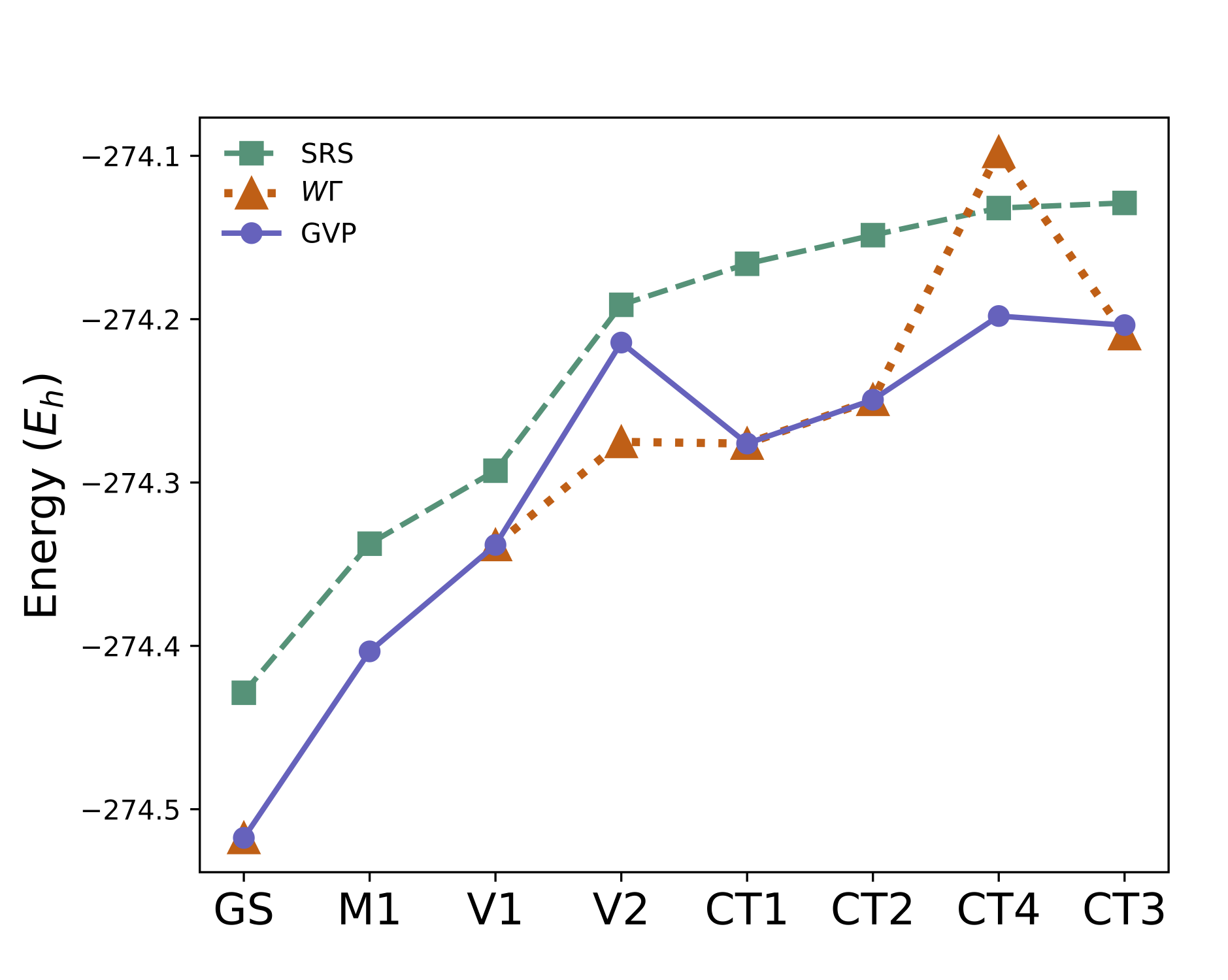}
    \caption{Energy ordering of first eight MgO ${}^1A_1$ states of the initial CASCI roots and after optimization with the $W\Gamma$ and GVP approaches.}
    \label{fig:MgO_Eordering}
\end{figure}

\begin{table}[htbp]
\caption{Energies (E$_h$) of the ${}^{1}A_{1}$ states in MgO,
listed from top to bottom in ascending order of the CASCI-LDA energies.
Labels (GS, M1, etc) are taken from a previous study. \cite{2step} }
\small
\begin{tabular}{c c c c c }
\hline\hline
State & Label & CASCI & $W\Gamma$ & GVP \\ \hline
$1^1A_1$ & GS & $\quad$-274.42869956 & $\quad$-274.51755503 & $\quad$-274.51755511 \\
$2^1A_1$ & M1 & $\quad$-274.33744776 & $\quad$-- & $\quad$-274.40336697 \\
$3^1A_1$ & V1 & $\quad$-274.29276479 & $\quad$-274.33820474 & $\quad$-274.33820504 \\
$4^1A_1$ & V2 & $\quad$-274.19120544 & $\quad$-274.27510790 & $\quad$-274.21432010 \\
$5^1A_1$ & CT1 & $\quad$-274.16609490 & $\quad$-274.27614863 & $\quad$-274.27614914 \\
$6^1A_1$ & CT2 & $\quad$-274.14857162 & $\quad$-274.24932368 & $\quad$-274.24932934 \\
$7^1A_1$ & CT4 & $\quad$-274.13197362 & $\quad$-274.09760158 & $\quad$-274.19806809 \\
$8^1A_1$ & CT3 & $\quad$-274.12884711 & $\quad$-274.20910669 & $\quad$-274.20364194 \\  \hline\hline
\end{tabular}
\label{tab:MgO_energies}

\end{table}

\renewcommand{\thesection}{\Roman{section}}
\setcounter{section}{6}
\section{References}
\bibliographystyle{achemso}
\bibliography{main}

\end{document}